\documentclass[9pt,twocolumn,twoside]{osajnl}
\journal{ao}
\usepackage{siunitx}
\setboolean{shortarticle}{false} 
\ifthenelse{\boolean{shortarticle}}{\colorlet{color2}{color2b}}{\colorlet{color2}{color2}}

\title{Spectral response of Bragg gratings in multimode polymer waveguides}

\author[1]{Aditya Bhuvaneshwaran}
\author[1]{Stanislav Sherman}
\author[1]{Hans Zappe}

\affil[1]{Gisela and Erwin Sick Chair of Micro-optics, Department of Microsystems Engineering, University of Freiburg, Germany}

\affil[1]{Corresponding author: aditya.bhuvaneshwaran@gmail.com}
\affil[1]{Corresponding author: zappe@imtek.uni-freiburg.de}

\dates{Compiled \today}

\ociscodes{(50.2770) Gratings; (130.6010) Sensors, (130.5460) Polymer waveguides; (060.3735) Fiber Bragg gratings; (030.4070) Modes. }

\doi{© 2017 Optical Society of America.]. One print or electronic copy may be made for personal use only. Systematic reproduction and distribution, duplication of any material in this paper for a fee or for commercial purposes, or modifications of the content of this paper are prohibited.
\url{http://dx.doi.org/10.1364/ao.XX.XXXXXX}}

\begin{abstract}
A means to calculate the multimodal spectral response of Bragg gratings in general non-circular multimode waveguides is proposed. To illustrate the power of the technique, the spectra of two Bragg temperature sensors are numerically calculated in which coupling between 100 modes considered. It is shown how the Bragg wavelength in multimode Bragg grating waveguides is affected by the number of modes and energy distribution among them. Good matching of the simulated spectrum of a multimode Bragg grating on a planar inverted rib waveguide to the measured spectrum is seen.
\end{abstract}

\setboolean{displaycopyright}{true}

\begin{document}

\maketitle
\thispagestyle{fancy}
\ifthenelse{\boolean{shortarticle}}{\ifthenelse{\boolean{singlecolumn}}{\abscontentformatted}{\abscontent}}{}

\section{Introduction}
Fiber Bragg gratings (FBGs) are extensively used for distributed measurement of physical parameters such as temperature \cite{Jung.1999} or strain \cite{Lim.2002} and as a result have been extensively studied in both the single-mode and multi-mode regimes. Kogelnik \cite{Kogelnik.1969} used coupled mode theory (CMT) to derive the spectral response of an electric field interacting with a refractive index perturbation; Erdogan \cite{Erdogan.1997} extended this approach and presented coupled first order differential equations describing the interaction of multiple modes in circular waveguides. Uniform and non-uniform gratings can be modeled using the transfer matrix method \cite{Little.1995} (TMM). Using these techniques, it is thus possible to obtain analytic expressions for the Bragg wavelength $\lambda_b$ as a function of temperature \cite{Nam.2006} for circular waveguides and thus determine the temperature sensitivity of a FBG sensor.

\begin{figure}[t]
\centering
\fbox{\includegraphics[width=0.75\columnwidth]{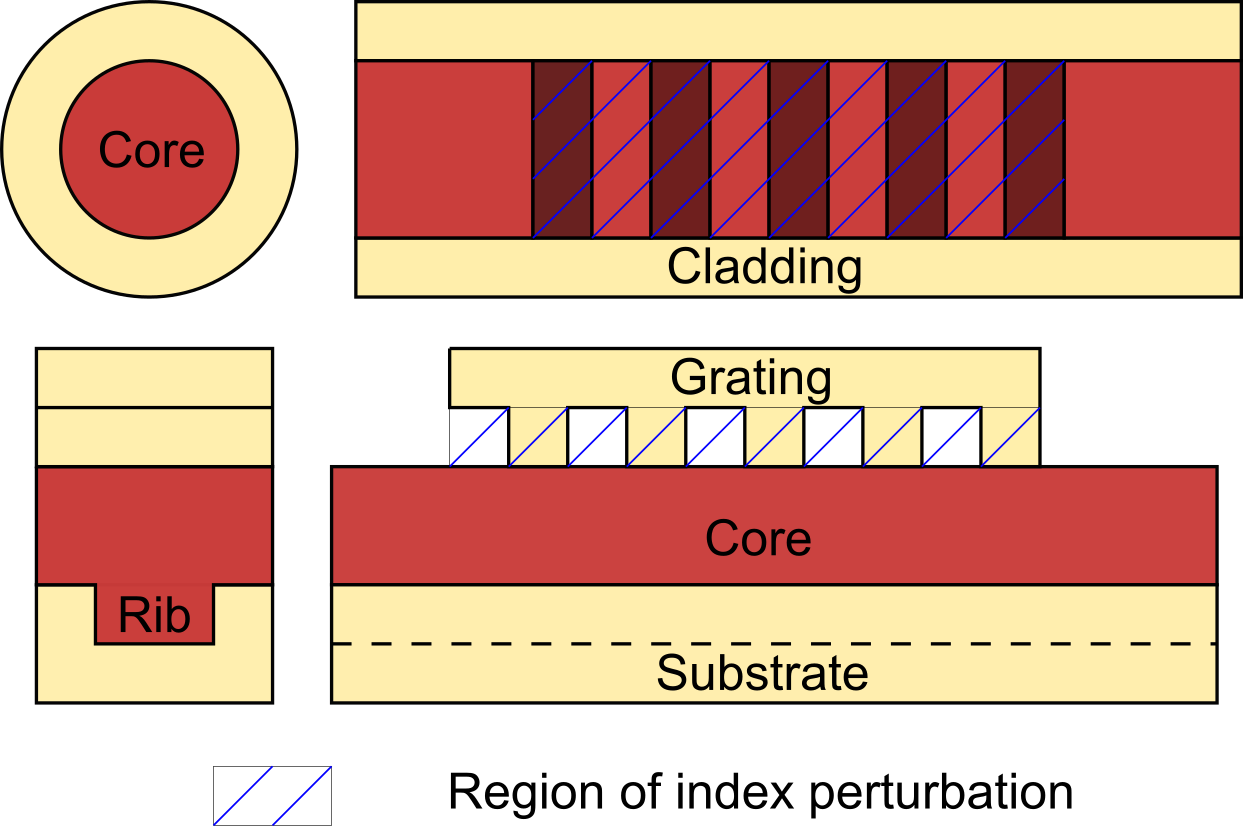}}
\caption{Fiber Bragg gratings with circular (FBG, top) and waveguide Bragg gratings with rectangular (RBG, bottom) cross-sections.}
\label{fig:GratingType}
\end{figure}

In the past decade, planar waveguides with Bragg gratings have become of increasing interest, since they can be fabricated using planar manufacturing techniques such as hot embossing \cite{Huang.2013} and nano imprinting \cite{Nam.2006}, hence avoiding the necessity of expensive techniques such as UV interference lithography.  As seen in Figure~\ref{fig:GratingType}, in contrast to fibers, the core cross section of planar waveguides is typically non-circular, such as the rectangular Bragg grating (RBG) based on an inverted rib waveguide shown in the figure.

Understanding the effects of non-circular cross sections and index perturbations in the multimode domain on the spectral response helps to provide insight into optical behavior as well as manufacturing requirements for planar waveguides used as Bragg sensors. However, four approximations used in the derivation of analytical expressions for the spectral response of FBGs need to be re-examined to compute the spectral response of RBGs:

\begin{enumerate}
\item A 3D model of FBGs can be simplified using cylindrical symmetry to a 1D problem \cite{Haus.1987}; however, RBGs can be reduced to at-most a 2D model.
\item In the case of 2D models, the intensity profile of the fundamental mode propagating in a FBG can be well approximated by a Gaussian function \cite{Kim.2007}; in RBGs, the higher modes have to be described with Bessel functions \cite{Tsai.2012}.
\item The refractive index profile of FBGs is found over the complete cross section of the core, as seen in Figure~\ref{fig:GratingType}; for RBGs, it changes only in a portion of the cross section and, in addition, the index perturbation can be in either the core, the cap or the substrate.
\item The magnitude of energy exchange between forward and backward traveling modes in a fiber is represented by the coupling coefficient $\kappa$. Given assumptions two and three above, and a grating with index perturbation $\delta n_{eff}$, $\kappa$ is often approximated as $2\pi\frac{\delta n_{eff}}{\lambda}$ \cite{Erdogan.1997} for single mode FBGs for light with wavelength $\lambda$. In RBGs, $\kappa$ has multiple values for every possible mode combination and thus must be numerically computed.
\end{enumerate}

We thus propose here modifications to these approximations to derive the multimodal spectral response of a grating placed on top of the core of an inverted rib waveguide; such a grating is referred to as a \textit{surface Bragg grating}. A commercial computation tool is used to perform simulations and simulated spectral responses are compared with actual measurements. We thus derive a means to easily compute the spectral response of a multi-mode RBG and use this to derive its response to temperature changes when used as a sensor.

\section{Theory}\label{sect:theory}
\begin{figure}[t]
	\centering
	\begingroup%
	\makeatletter%
	\setlength{\unitlength}{0.45\columnwidth}
	\makeatother%
	\begin{picture}(1,1.00000023)%
	\put(0,0){\fbox{\includegraphics[width=\unitlength,page=1]{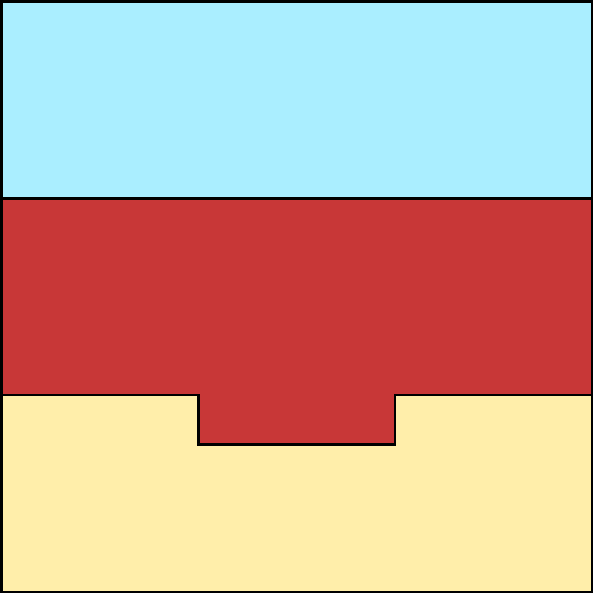}}}%
	\put(0.51909067,0.7986683){\color[rgb]{0,0,0}\makebox(0,0)[b]{\smash{Cap $n_{\circ}$}}}%
	\put(0.50981425,0.4716249){\color[rgb]{0,0,0}\makebox(0,0)[b]{\smash{Core $n_{g}$ }}}%
	\put(0.50476381,0.11129452){\color[rgb]{0,0,0}\makebox(0,0)[b]{\smash{Substrate $n_{s}$}}}%
	\put(0.53109567,0.27518026){\color[rgb]{0,0,0}\makebox(0,0)[b]{\smash{Rib $n_{g}$ }}}%
	\put(0,0){\fbox{\includegraphics[width=\unitlength,page=2]{Pics_Sensor_FrontView_01_Waveguide.pdf}}}%
	\put(0.08677833,0.2002982){\color[rgb]{0,0,0}\makebox(0,0)[b]{\smash{y}}}%
	\put(0.1979683,0.06695455){\color[rgb]{0,0,0}\makebox(0,0)[b]{\smash{x}}}%
	\put(0.04088154,0.06661794){\color[rgb]{0,0,0}\makebox(0,0)[b]{\smash{z}}}%
	\end{picture}%
	\endgroup%
	\hfill
	\fbox{\includegraphics[width=0.45\columnwidth,keepaspectratio=true]{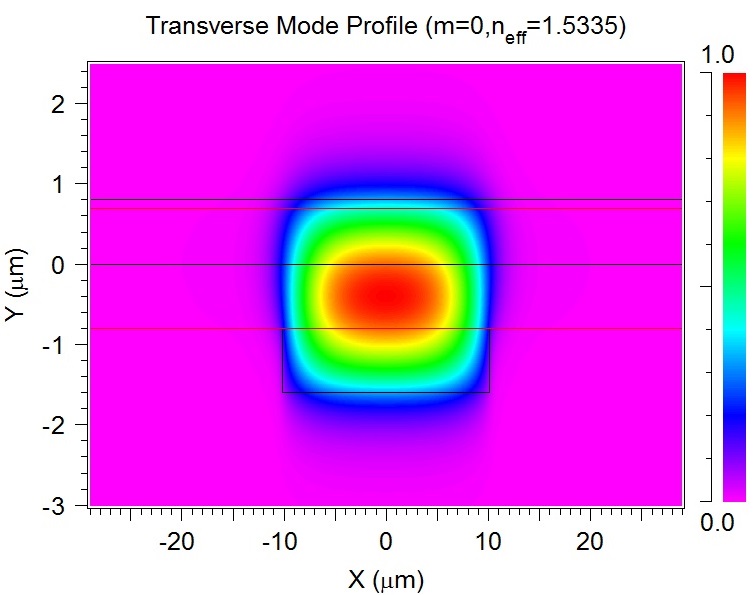}}
	\caption{Cross section of an inverted rib waveguide and the electric field distribution of the fundamental mode propagating in it.}
	\label{fig:CrossWav}
\end{figure}

We begin by proposing a general 3D expression describing the electric field, $\vec{E}_w$, in a RBG  supporting multiple modes, whose unit mode profiles $\vec{e_{t,j}}$ are 2D non-Gaussian functions. The interaction of this field with the grating is analysed and coupled first-order differential equations are formulated, overcoming assumptions 3 and 4 above. These equations are numerically solved and an equation describing the multimodal spectral response of a surface Bragg grating, as shown in Figure~\ref{fig:CrossWav}, is derived.

\subsection{Multimode propagation}\label{sect:MMProp}
\begin{figure}[t]
\centering
\begingroup%
\makeatletter%
\setlength{\unitlength}{\columnwidth}
\makeatother%
\begin{picture}(1,0.29251514)%
\put(0,0){\fbox{\includegraphics[width=\unitlength,page=1]{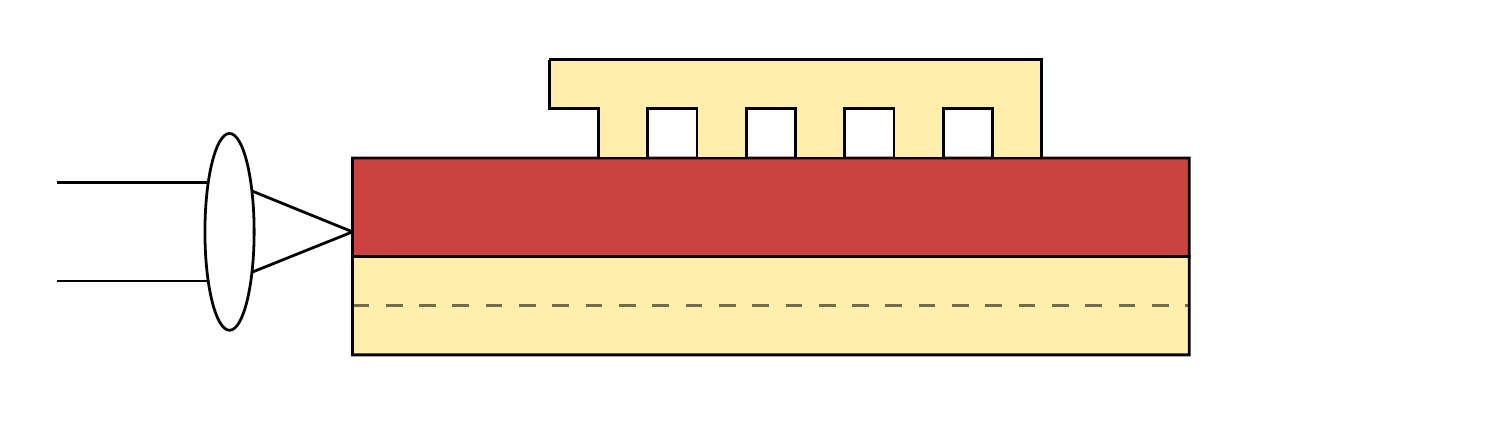}}}%
\put(0.07778926,0.12653073){\color[rgb]{0,0,0}\makebox(0,0)[b]{\smash{$I(\lambda)$}}}%
\put(0,0){\includegraphics[width=\unitlength,page=2]{Pics_ElectricField.pdf}}%
\put(0.15446244,0.02616405){\color[rgb]{0,0,0}\makebox(0,0)[b]{\smash{$\zeta_{c}$}}}%
\put(0,0){\fbox{\includegraphics[width=\unitlength,page=3]{Pics_ElectricField.pdf}}}%
\put(0.3178534,0.14425723){\color[rgb]{0,0,0}\makebox(0,0)[b]{\smash{$\vec{E_w}$}}}%
\put(0.42240165,0.14425723){\color[rgb]{0,1,0}\makebox(0,0)[b]{\smash{$\vec{E_r}(z)$}}}%
\put(0.608402,0.14425723){\color[rgb]{0,0,1}\makebox(0,0)[b]{\smash{$\vec{E_t}(z)$}}}%
\put(0,0){\fbox{\includegraphics[width=\unitlength,page=4]{Pics_ElectricField.pdf}}}%
\put(0.44887023,0.07031056){\color[rgb]{0,0,0}\makebox(0,0)[b]{\smash{$z$}}}%
\put(0.14067474,0.23254238){\color[rgb]{0,0,0}\makebox(0,0)[b]{\smash{Incoupling lens}}}%
\put(0.23732001,0.00589955){\color[rgb]{0,0,0}\makebox(0,0)[b]{\smash{$\zeta_{e}$}}}%
\put(1.03936282,0.14607238){\color[rgb]{0,0,0}\makebox(0,0)[b]{\smash{}}}%
\put(0.79857834,0.01415825){\color[rgb]{0,0,0}\makebox(0,0)[b]{\smash{$\zeta_{o}$}}}%
\put(0.8858232,0.03341861){\color[rgb]{0,0,0}\makebox(0,0)[b]{\smash{$\zeta_{s}$}}}%
\put(0.86775479,0.27210624){\color[rgb]{0,0,0}\makebox(0,0)[b]{\smash{Coupling lens to }}}%
\put(0.8623715,0.22968157){\color[rgb]{1,0.4,0}\makebox(0,0)[b]{\smash{Spectrometer}}}%
\end{picture}%
\endgroup%
\caption{Interaction of $E_w$ with the grating and the losses occurring at each stage.}
\label{fig:CoupleModeSolutions}
\end{figure}

Laser diodes; light emitting diodes; and super-luminescent diodes are commonly used as light sources for fiber and waveguide sensors, and their intensity profiles can be assumed to a good approximation to be a Gaussian function \cite{Kim.2007} (with peak intensity $I_\circ$, peak wavelength $\lambda_\circ$ and FWHM $\sigma$). The incoupled light $I(\lambda)$ experiences coupling losses $\zeta_{c}$ due to Fresnel reflections when it is incident on the waveguide facet as shown in Figure~\ref{fig:CoupleModeSolutions}. The electric field $\vec{E_e}$ at the facet then excites a wave in the waveguide $\vec{E_w}$ given by
\begin{align}\label{eq:Lightsource}
|\vec{E_w}(\lambda)|^2
& = \zeta_{e} |\vec{E_e}(\lambda)|^2 \\ \nonumber
& = \zeta_{e} \zeta_{c} \cdot I(\lambda) \\ \nonumber
& = \zeta_{e} \zeta_{c} \cdot \frac{I_\circ}{\sqrt{2 \pi \sigma}} \exp \left(\frac{-(\lambda-\lambda_\circ)^2}{2\sigma^2}\right)
\end{align}
where $\zeta_{e}$ represents the excitation loss.

The parameters of Equation \ref{eq:Lightsource} are shown graphically in Figure~\ref{fig:CoupleModeSolutions} for a surface Bragg grating with $p$ modes $\vec{A_j}$ propagating toward the grating and $p$ modes $\vec{B_j}$ reflected from the grating. The forward propagating electric field $\vec{E_w}$ is the superposition of $p$ modes and is formulated as
\begin{equation}
\vec{E_w}(\lambda, z) = \zeta_{e}\sum_{0}^{p-1}|\vec{E_e}(\lambda)| \xi_j \vec{A_j}(z) .
\end{equation}
In this expression, $\vec{E_w}$ is a function of wavelength $\lambda$ and distance $z$ in the longitudinal direction. The origin is placed at the start of the grating. Each mode $j$ is a distinct wave, represented by $\vec{A_j}$, namely
\begin{subequations}
\begin{equation}
\vec{A_j}(z) = e^{i(\beta_j z - \omega t)}\cdot \vec{e}_{t,j}(x,y) \cdot e^{-\alpha z}
\end{equation}
\begin{equation}
|\vec{A_j}(z=0)| = 1
\end{equation}
\end{subequations}
and each mode is defined as having unit power. $\vec{e}_{t,j}(x,y)$ is the transverse mode profile of the wave as shown in Figure~\ref{fig:CrossWav}.

The distribution of energy among modes \cite{Wood.1984} \cite{Tsekrekos.2007} is described by the modal transfer function \cite{Olivero.2010}, where $\xi_j$ is the fraction of power in mode $j$. For a wave with unit power, the modal transfer function is defined as
\begin{equation}
\sum_{0}^{p-1}\xi_j = 1 .
\end{equation}
We discuss this aspect of the energy distribution in the modes further in Section \ref{sect:Design}\ref{sect:ModTransFunc} as to how the spectral response of a multimode RBG can be influenced by it.

The Bragg grating now reflects a fraction of the incident wave $\vec{E_w}$. Each reflected mode $B_j(z)$ can be re-written as the product of incident wave $\vec{A_j}(0)$ times reflectivity of the mode in the grating $r_j(z)$. The reflected wave $\vec{E_r}$ is then equal to the sum of reflected modes $\vec{B_j}$ as given by
\begin{equation}
\vec{E_r}(z) = \sum_{0}^{p-1}\vec{B_j}(z) = |\vec{E_w}| \zeta_o \zeta_s \left(\sum_{0}^{p-1} \xi_i \vec{A_j}(0) \cdot r_j(z) \right) .
\end{equation}

The outcoupling efficiency $\zeta_o$ and efficiency of incoupling into spectrometer $\zeta_s$ are assumed to be 100\%. Thus, the transmitted light $\vec{E_t}$ as a function of $z$ is given by
\begin{equation}
\vec{E_t}(z) = \sum_{0}^{p-1}\vec{A_j}(z) = |\vec{E_w}| \zeta_o \zeta_s \left(\sum_{0}^{p-1} \xi_i \vec{A_j}(0) \cdot t_j(z) \right) ,
\end{equation}

where $r_j$ and $t_j$ are the reflectivity and transmissivity of each mode.

\subsection{Coupled mode theory}
Coupled mode theory is used to calculate $r_j$ and $t_j$; full derivation of CMT has been presented previously \cite{Haus.1987} \cite{Little.1995} \cite{Yariv.1973}. The electric field of the jth mode in the steady state, $E_j(z)$ can be described as
\begin{equation}
E_j(z) = (A_j(z)e^{i\beta_j z} + B_j(z)e^{-i\beta_j z} ) \cdot e^{-i\omega t} \cdot \vec{e_t}(x,y) .
\end{equation}
Based on Erdogan's description \cite{Erdogan.1997} of multimode waveguides, the magnitude of modes $A_j$ and $B_j$ evolve along $z$ as
\begin{subequations}
\begin{align} \label{eq:A}
\frac{\partial A_j}{\partial z} =
&\textrm{ }i \sum_{k = 0}^{p-1} \left( \kappa^t_{kj} + \kappa^l_{kj} \right) A_k e^{\left[ i\left( \beta_k - \beta_j \right) z \right]} \\ \nonumber
& + i \sum_{k = 0}^{p-1} \left( \kappa^t_{kj} - \kappa^l_{kj} \right) B_k e^{\left[ -i\left( \beta_k + \beta_j \right) z \right]}
\end{align}
\begin{align}\label{eq:B}
\frac{\partial B_j}{\partial z} =
& - i \sum_{k = 0}^{p-1} \left( \kappa^t_{kj} - \kappa^l_{kj} \right) A_k e^{\left[ i\left( \beta_k + \beta_j \right) z \right]} \\ \nonumber
& - i \sum_{k = 0}^{p-1} \left( \kappa^t_{kj} + \kappa^l_{kj} \right) B_k e^{\left[ -i\left( \beta_k - \beta_j \right) z \right]} .
\end{align}
\end{subequations}

Coupling coefficients \cite{Snyder.1972} $\kappa^t_{k,j}$ and $\kappa^l_{k,j}$ quantify the fraction of energy transferred from mode $j$ to each of the other modes $k$. They can be described in terms of the electric field mode profiles $\vec{e_{t,j}}$ and $\vec{e_{t,k}}$ of $\beta_j$ and $\beta_k$ as
\begin{subequations}
\begin{equation}\label{eq:kappat}
\kappa^t_{kj} = \frac{\omega }{2}
\frac{\iint(\epsilon - \epsilon_\circ) \overrightarrow{e_{t,k}}\cdot \overrightarrow{e_{t,j}}\cdot dx dy }
{\iint \overrightarrow{e_{t,j}} \times \overrightarrow{h_{t,j}}\cdot dx dy} 
\end{equation}
\begin{equation}
\kappa^l_{kj} = \frac{\omega }{2}
\frac{\iint\frac{
\epsilon_\circ}{\epsilon} (\epsilon - \epsilon_\circ)
\cdot
(\overrightarrow{e_{z,k}}\cdot \overrightarrow{e_{z,j}})
\cdot dx dy}
{\iint \overrightarrow{e_{t,j}} \times \overrightarrow{h_{t,j}}\cdot dx dy}
\end{equation}
\end{subequations}
where $\vec{h_{t,j}}$ is the magnetic field mode profile of $\beta_j$ and $\omega$ is $\frac{2\pi c}{\lambda}$.

For the case of modes in waveguides, $\kappa^l_{j,k}$ is much smaller than $\kappa^t_{j,k}$ \cite{Erdogan.1997}. Accordingly, $\kappa$ can be approximated as
\begin{subequations}
\begin{equation}
\kappa^t_{kj} + \kappa^l_{kj} \approx \kappa^t_{kj} = \kappa_{kj} 
\end{equation}
\begin{equation}
\kappa^t_{kj} - \kappa^l_{kj} \approx \kappa^t_{kj} = \kappa_{kj} .
\end{equation}
\end{subequations}

\begin{figure}[t]
\centering
\def\svgwidth{\columnwidth}
\fbox{\includegraphics[width=\columnwidth,keepaspectratio=true]{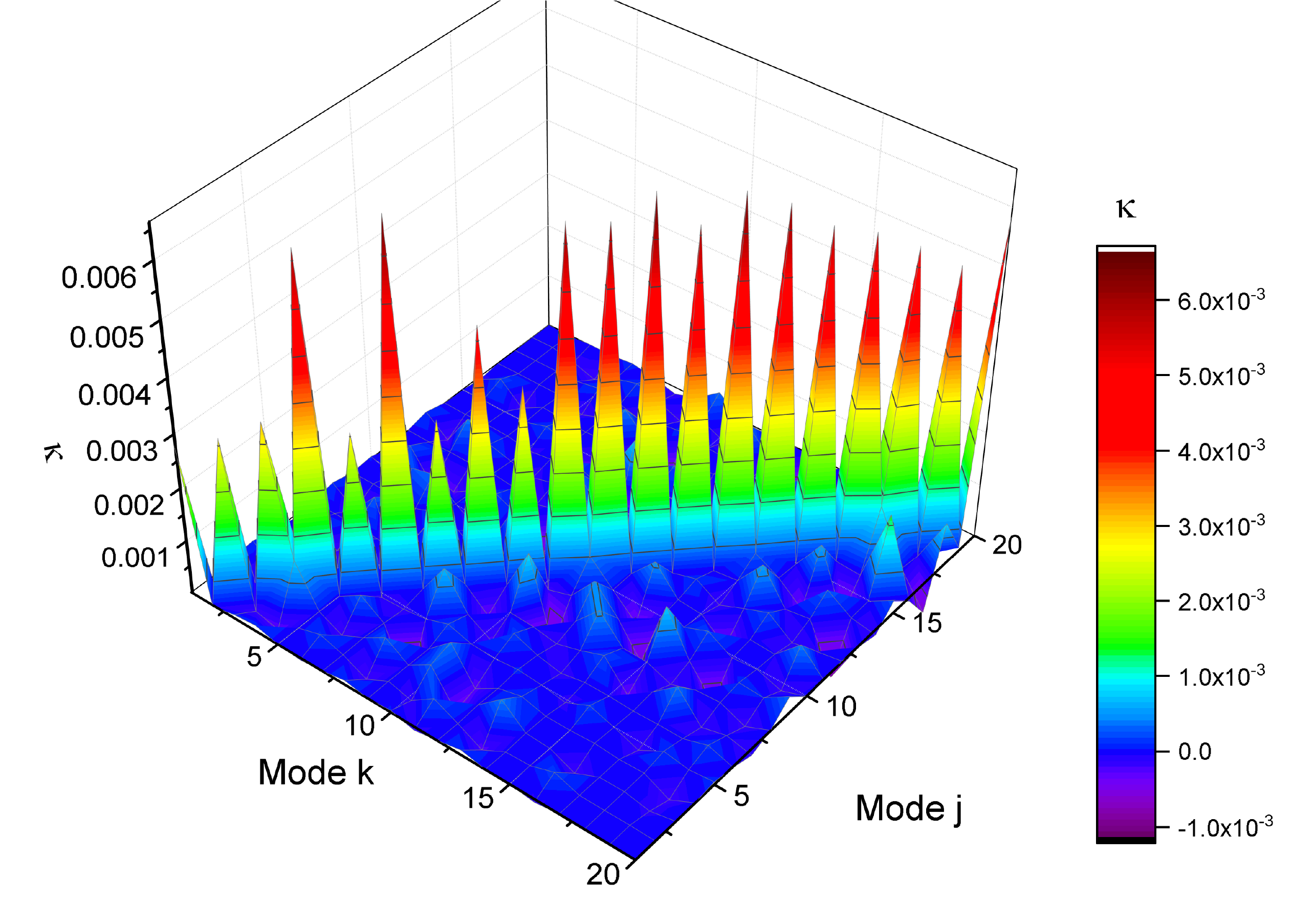}}
\caption{3D plot showing relative magnitudes of $\kappa_j$. Intra coupling coefficient $\kappa_{k\neq j,j}$ dominates inter coupling coefficients $\kappa_{k\neq j,j}$ }
\label{fig:Kappa_calc}
\end{figure}

The coupling of mode $j$ to itself is called intra-mode coupling $\kappa_{k=j}$ and from mode $j$ to $k$ is called inter-mode coupling $\kappa_{k\neq j}$. When $\kappa$ is computed (Appendix A.), we find that intra-modal coupling dominates (Figure~\ref{fig:Kappa_calc}), so that

\begin{subequations}
\begin{align}
\label{eq:kappapprox}
\kappa_{k=j} &\gg \kappa_{k\neq j} \\
\kappa_{k=j} &= \kappa_{j} .
\end{align}
\end{subequations}

Using Equation \ref{eq:kappapprox}, \ref{eq:A} and \ref{eq:B} then take the form
\begin{subequations}
\begin{equation} \label{eq:A2}
\frac{\partial A_j}{\partial z} = i \kappa_j A_j e^{ -2i\beta_j z } + i \kappa_j B_j
\end{equation}
\begin{equation}\label{eq:B2}
\frac{\partial B_j}{\partial z} = - i \kappa_j A_j - i \kappa_j B_j e^{ 2i\beta_j z } .
\end{equation}
\end{subequations}

\subsection{Transfer matrix method}
The transfer matrix method \cite{Little.1995} can now be used to solve these coupled differential equations numerically \cite{Kim.2006b}. Equations \ref{eq:A2} and \ref{eq:B2} can be expressed in matrices as functions of distance $z$ along the grating as
\begin{equation}\label{eq:Matrix}
\frac{\partial \begin{bmatrix}
A_j\\
B_j
\end{bmatrix}}{\partial z} = C(z) \begin{bmatrix}
A_j\\
B_j
\end{bmatrix} .
\end{equation}
The coupling matrix $C$ is composed of coupling coefficients, namely
\begin{equation}\label{eq:Matrix7}
C(z) = \begin{bmatrix}
i\kappa_j e^{ -i2\beta_j z} & i\kappa_j \\
-i\kappa_j & -i\kappa_j e^{ i 2\beta_j z }
\end{bmatrix} .
\end{equation}
The transfer matrix $F$ models the energy exchange between modes and is calculated from the coupling matrix $C$ using the TMM. Numerical computation of $F$ can be significantly accelerated using eigen-decomposition \cite{Huang.1994}. The solution to Equation \ref{eq:Matrix} can be written in terms of the transfer matrix $F$ as
\begin{equation}\label{eq:Matrix8_2}
\begin{bmatrix}
A_j(z=0)\\
B_j(z=0)
\end{bmatrix} = F \begin{bmatrix}
A_j(z=L)\\
B_j(z=L)
\end{bmatrix} .
\end{equation}

Equation \ref{eq:Matrix8_2} has four unknowns and two equations. By definition, the magnitude of $A_j$ at the start of the grating ($z = 0$) is unity and the magnitude of $B_j$ at the end of the grating ($z = L$) is zero,
\begin{subequations}
\begin{equation}\label{eq:BoundaryCondition1}
|A_j(z=0)| = 1
\end{equation}
\begin{equation}\label{eq:BoundaryCondition2}
|B_j(z=L)| = 0 ,
\end{equation}
\end{subequations}
so that substituting these boundary conditions in Equation \ref{eq:Matrix8_2}, yields
\begin{equation}\label{eq:Matrix9}
\begin{bmatrix}
1\\
B_j(0)
\end{bmatrix} = F \begin{bmatrix}
A_j(L)\\
0
\end{bmatrix} .
\end{equation}

The transfer matrix contains numerical values from which reflectivity $r_j$ and transmissivity $t_j$ can be expressed as
\begin{subequations}
\begin{equation}
r_j = B_j(0) 
\end{equation}
\begin{equation}
t_j = A_j(L) .
\end{equation}
\end{subequations}

\subsection{Multimode Spectrum}
To determine the multimode spectrum, each reflected mode $B_j(z)$ is calculated for each incident mode $A_j(z)$. The multimode reflectivity $r_m$ can be defined as the ratio of reflected intensity to incident intensity of the wave as
\begin{equation}\label{eq:Start2}
r_m(z, \lambda) = \left|\frac{\vec{E_r}(\lambda)}{\vec{E_w}(\lambda)}\right|^2 = \left( \sum_{0}^{p-1} \xi_j |\vec{A_j}(0)| \cdot r_j(L, \lambda)\right)^2 .
\end{equation}
Since $|A_i(0)| = 1$, we have
\begin{equation}\label{eq:reflectivity}
r_m(0, \lambda) = \left(\sum_{0}^{p-1}r_i(\lambda) \xi_i\right)^2 .
\end{equation}

Similarly, an equation for $z=L$ can be written for Equation \ref{eq:Start2} and we can obtain overall transmissivity as
\begin{equation}\label{eq:transmissivity1}
t_m(z, \lambda) = \left|\frac{\vec{E_t}(\lambda)}{\vec{E_w}(\lambda)}\right|^2 = \left(\sum_{0}^{p-1}\xi_j |\vec{A_j}(0)| \cdot t_j(L, \lambda)\right)^2 
\end{equation}
such that
\begin{equation}\label{eq:transmissivity}
t_m(L, \lambda) = \left(\sum_{0}^{p-1}t_j(z,\lambda) \xi_j\right)^2 .
\end{equation}
Thus, the spectra of individual modes are combined to obtain Equations \ref{eq:reflectivity} and \ref{eq:transmissivity} which the describe the spectral response of a RBG.

\subsection{Temperature dependence}
Since one important application of Bragg gratings is for temperature sensing, and that application will be used as an example here, we consider the expected change in the spectrum as a function of temperature. To that end, the physical length of the waveguide varies as a function of temperature (T) as
\begin{equation}\label{eq:temp_l}
L(\Delta T) = L_\circ (1 + \alpha \Delta T)
\end{equation}
where $\alpha$ is the linear thermal expansion coefficient and $\Delta T = T_\circ - T$ and $T_\circ = 273 \textrm{K}$. The material refractive index of the cap, core and substrate layers, $n_c, n_g, n_s$, vary as
\begin{equation}\label{eq:temp_n}
n(\Delta T) = n_\circ + N^\prime \Delta T
\end{equation}
where $N^\prime$ is the thermo-optic coefficient of the material.

Equations \ref{eq:temp_l} and \ref{eq:temp_n} are substituted into Equation \ref{eq:Matrix7} \cite{Yamada.1987} . Thus peak reflectivity $r_m$ is then found as a function of $T$.

The sensitivity of FBGs to changes in temperature, $S_s$, can be expressed using the analytic expression \cite{Jung.1999}
\begin{align}
S_s
&= \frac{\partial \lambda_b}{\partial T}\\
&= 2 \frac{\partial n_{eff} }{\partial T} \Lambda + 2 n_{eff} \frac{\partial \Lambda}{\partial T}\\
&= 2\left( N^\prime \Lambda + \alpha n_{eff} \right) .
\end{align}
As an alternative, we may employ the multimode reflectivity $r_m(\lambda)$ for planar surface Bragg gratings, which is a numerically computed value and is calculated as a function of $\lambda$; the wavelength with the highest reflectivity is denoted as the Bragg wavelength, $\lambda_b$. Even though $\lambda_b$ is not an analytic expression, a more general relationship,
\begin{equation}\label{eq:Sensitivity}
S_m = \frac{\partial \lambda_b}{\partial T} ,
\end{equation}
may also be used to define sensitivity.

\section{Fabrication}
To define the structures which will be subject to analysis by simulation in Section \ref{sect:Design}, we briefly consider the fabrication techniques used for the rectangular Bragg grating structures employing a planar surface Bragg gratings which will then provide the experimental results of Section \ref{sect:Meas}. 

\begin{figure}[t]
\centering
\begin{minipage}{.45\columnwidth}
\begingroup%
\makeatletter%
\setlength{\unitlength}{\textwidth}
\makeatother%
\begin{picture}(1,1.50000003)%
\put(0,0){\fbox{\includegraphics[width=\unitlength,page=1]{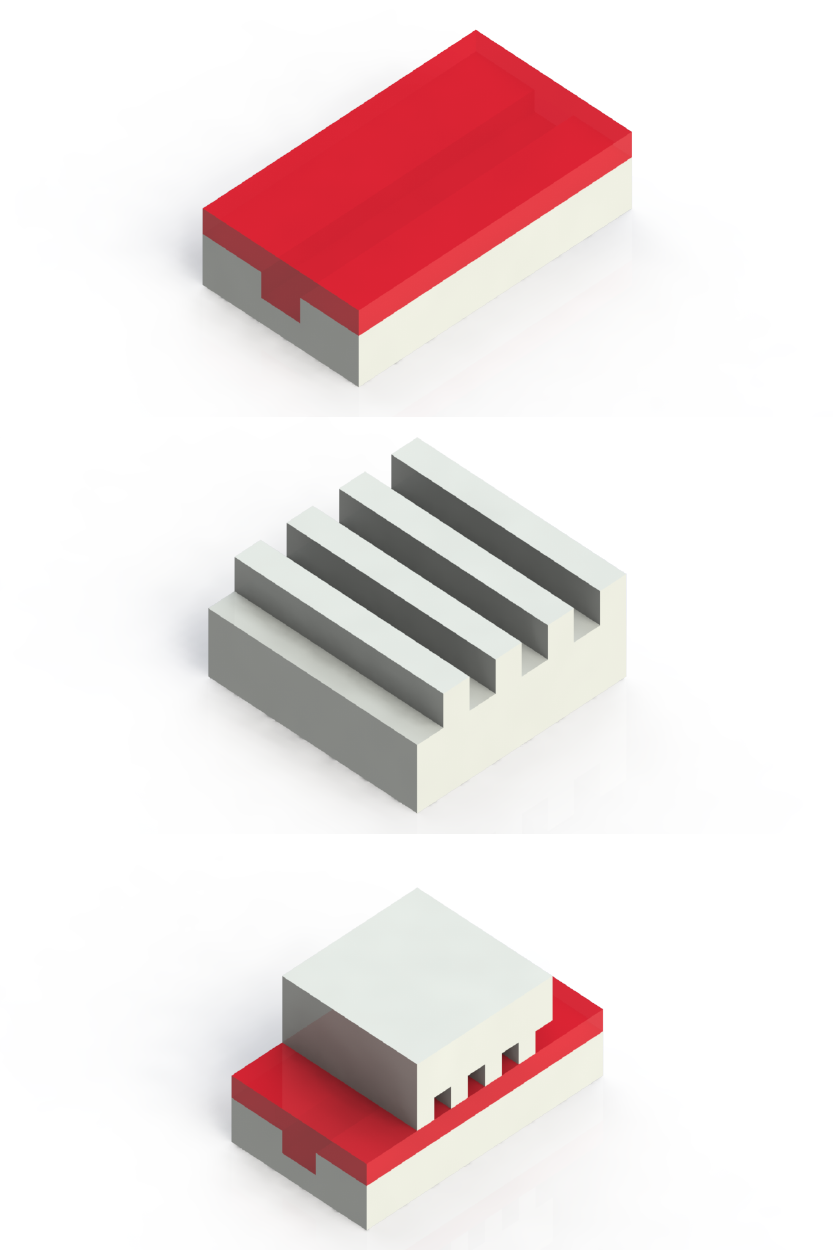}}}%
\put(0.12790839,1.17660322){\color[rgb]{0,0,0}\makebox(0,0)[b]{\smash{Step 1}}}%
\put(0.13505124,0.67660324){\color[rgb]{0,0,0}\makebox(0,0)[b]{\smash{Step 2}}}%
\put(0.12790839,0.17660319){\color[rgb]{0,0,0}\makebox(0,0)[b]{\smash{Step 3}}}%
\end{picture}%
\endgroup%
\caption{Cap Sensor}\label{fig:CapSensor}
 \end{minipage}\hfill
 \begin{minipage}{0.45\columnwidth}
\begingroup%
\makeatletter%
\setlength{\unitlength}{\textwidth}
\makeatother%
\begin{picture}(1,2.00000003)%
\put(0,0){\fbox{\includegraphics[width=\unitlength,page=1]{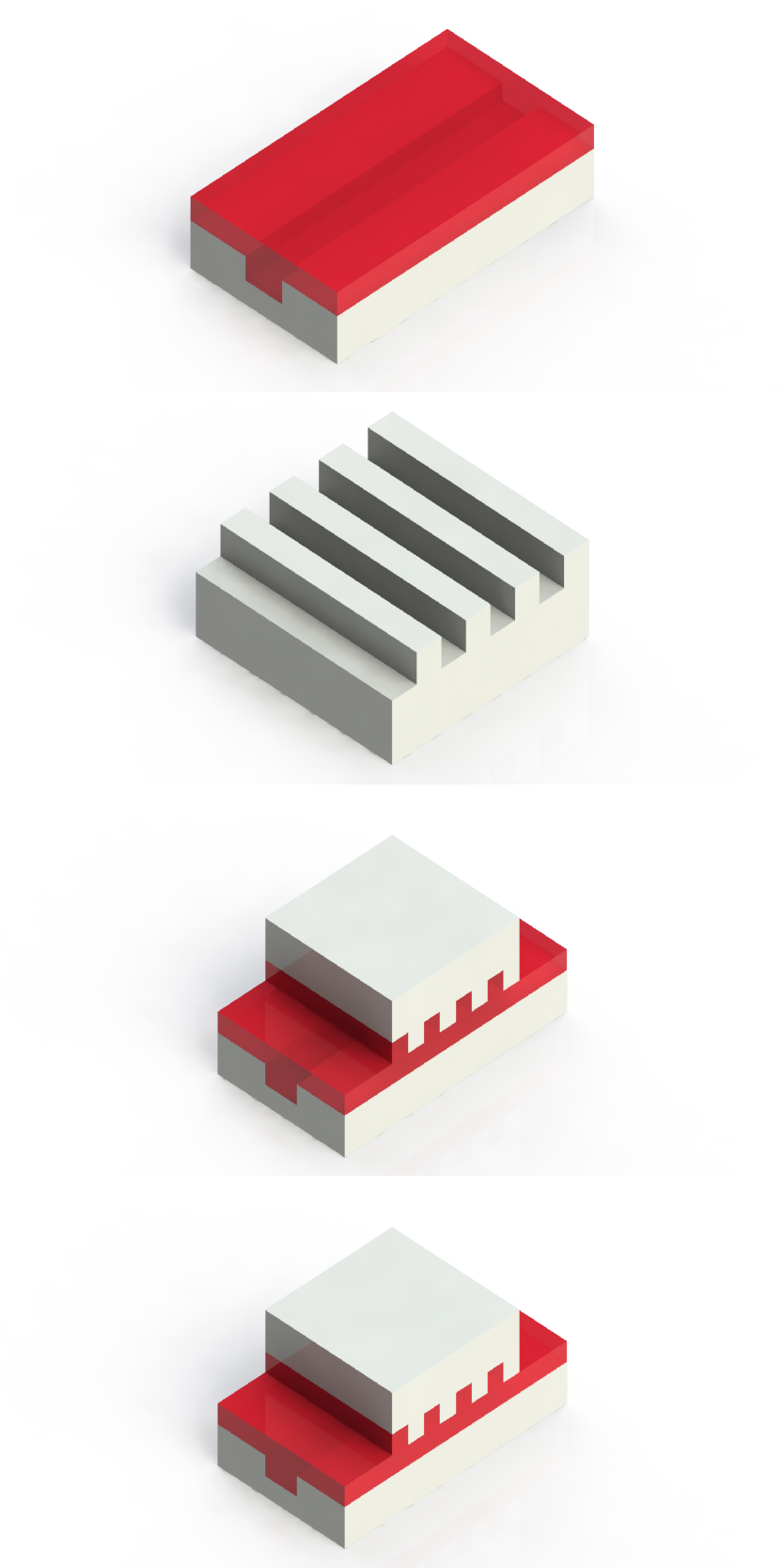}}}%
\put(0.13127838,1.67660322){\color[rgb]{0,0,0}\makebox(0,0)[b]{\smash{Step 1}}}%
\put(0.13127838,1.17660324){\color[rgb]{0,0,0}\makebox(0,0)[b]{\smash{Step 2}}}%
\put(0.13127838,0.67660319){\color[rgb]{0,0,0}\makebox(0,0)[b]{\smash{Step 3}}}%
\put(0.13127838,0.17660329){\color[rgb]{0,0,0}\makebox(0,0)[b]{\smash{Step 4}}}%
\end{picture}%
\endgroup%
\caption{Core Sensor}\label{fig:CoreSensor}
\end{minipage}
\end{figure}

The planar waveguide (Figure~\ref{fig:GratingType}) was realized using UGS45E, a mixture of $45\%$ ethylene glycol dimethyl acrylate and $55\%$ Syntholux, a commercially available UV-curable mixture based on an epoxy acrylate, by weight and poly(methyl meth-acrylate) (PMMA) as the substrate ($n_s$) \cite{Xiao.2015}. A hot embossed grating was defined using PMMA \cite{Sherman.2014} and allowed realization of two grating configurations, which we refer to as the cap sensor (Figure~\ref{fig:CapSensor}) and the core sensor (Figure~\ref{fig:CoreSensor}). The cap sensor is made by placing a grating on top of polymerised core layer whereas a core sensor is made by pressing the grating stamp into the unpolymerised (soft) core layer, followed by UV polymerisation.

\section{Design \& simulation}\label{sect:Design}
Based on the theoretical considerations above and the structure of the Bragg gratings which can be fabricated, CAD models of the cap and core sensors were defined using the commercial simulation software \emph{Rsoft}, in particular the module \emph{BeamPROP} which uses the beam propagation method \cite{Hadley.1992} \cite{vanRoey.1981} (BPM) to estimate mode profiles $\vec{e_t}$. The module \emph{GratingMOD} was used to solve Equations \ref{eq:A2} and \ref{eq:B2} numerically.

\subsection{Assumptions}
A number of simplifying assumptions for the simulations were made at the outset:
\begin{enumerate}
  \setlength\itemsep{0em}
\item Modal power redistribution due to scattering is neglected;
\item All features have zero deviations from the ideal;
\item All modes are uniformly TE polarised;
\item The polarisation is constant as the wave propagates;
\item Dispersion is assumed to be zero;
\item Scattering \& absorption losses are zero;
\item All thermal coefficients are linear; and
\item Loss coefficients $\zeta_c$, $\zeta_e$, $\zeta_o$ and $\zeta_s$ are assumed to be independent of $\lambda$.
\end{enumerate}
We discuss the implications of some of these assumptions on the results in Section \ref{sect:Meas}\ref{sect:Discussion}.

\subsection{Model input parameters}
\begin{figure}[t]
	\centering
	\begin{minipage}{.45\columnwidth}
		\centering
		\begingroup%
		\makeatletter%
		\setlength{\unitlength}{\textwidth}
		\makeatother%
		\begin{picture}(1,1.00000023)%
		\put(0,0){\fbox{\includegraphics[width=\unitlength,page=1]{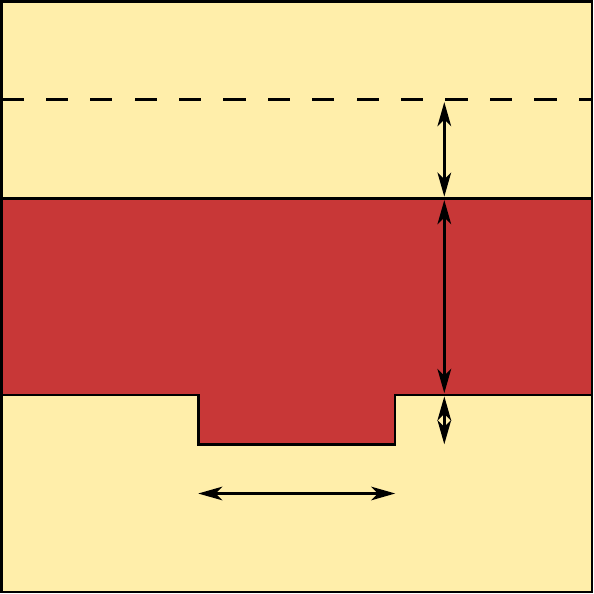}}}%
		\put(0.82704359,0.72400141){\color[rgb]{0,0,0}\makebox(0,0)[b]{\smash{ED}}}%
		\put(0.84347698,0.47659176){\color[rgb]{0,0,0}\makebox(0,0)[b]{\smash{$H_{c}$}}}%
		\put(0.83962875,0.26299371){\color[rgb]{0,0,0}\makebox(0,0)[b]{\smash{$H_{r}$}}}%
		\put(0.50000001,0.0969883){\color[rgb]{0,0,0}\makebox(0,0)[b]{\smash{$W_{r}$}}}%
		\put(0,0){\fbox{\includegraphics[width=\unitlength,page=2]{Pics_SimulationParameters_01_Geometric.pdf}}}%
		\put(0.1481922,0.76095195){\color[rgb]{0,0,0}\makebox(0,0)[b]{\smash{$H_c^2$}}}%
		\put(0.25117043,0.51839257){\color[rgb]{0,0,0}\makebox(0,0)[b]{\smash{$C_w$}}}%
		\put(0.16331605,0.13512241){\color[rgb]{0,0,0}\makebox(0,0)[b]{\smash{$H_s$}}}%
		\end{picture}
		\endgroup%
		\caption{Definition of the dimensional parameters used in the simulation.}
		\label{fig:Geometric}
	\end{minipage}\hfill
	\begin{minipage}{0.45\columnwidth}
		\centering
		\fbox{\includegraphics[width=\columnwidth,keepaspectratio=true]{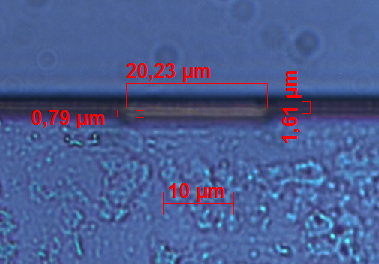}}
		\caption{Measured dimensions of a fabricated waveguide configured as a cap sensor}\label{fig:DesignDimensionsMeasurements}
	\end{minipage}
\end{figure}

Common design variables of both (cap and core) sensors are illustrated in Figure~\ref{fig:Geometric} and tabulated in Table \ref{tab:GeomParam}. The two sensors differ in the location of the Bragg index modulation and its magnitude. In both sensors, the index modulation has a square wave pattern and is not across the entire cross section of the waveguide. Structural details are illustrated in Appendix \ref{sect:appendixB} 


For the cap sensor, the propagating wave encounters two alternating materials (PMMA and air) in the cap, with an index modulation of $\delta n_{cap} = 1.488-1 = 0.488$. For the core sensor, the index perturbation is in the core and has a modulation of $\delta n_{core} = 1.543-1.488 = 0.055$.  The actual dimensions of the fabricated sensors (Figure \ref{fig:DesignDimensionsMeasurements}) were measured after fabrication and used to improve the accuracy of the simulations. The most relevant parameters are listed in Table \ref{tab:MeasGeomParam}. In addition, the material properties of the core and cladding layers were characterized experimentally and the measured values are given in Table \ref{tab:CharParam}.

For readers interested in replicating the simulations, parameters affecting numerical accuracy with which CMT \& BPM are computed using the finite element method (FEM) are listed in Table \ref{tab:SoftParam} of Appendix \ref{sect:appendixC}.

\begin{table}[H]
\centering
\caption{Design parameters} 
\begin{tabular}{lcc}
\hline Name & Symbol & Design \\ 
\hline Core height & $H_{c}$ & 2 $\mu \textrm{m}$ \\ 
Rib height & $H_{r}$ & 0.8 $\mu \textrm{m}$ \\ 
Rib width & $W_{r}$ & 20 $\mu \textrm{m}$ \\ 
Etch depth & ED & 110 nm \\ 
Duty cycle & DC & $50\%$ \\ 
Core Width & $C_w$ & 60 $\mu \textrm{m}$ \\ 
Cap Height & $H_c^2$ & 3 $\mu \textrm{m}$ \\ 
Substrate Height & $H_s$ & 4 $\mu \textrm{m}$ \\ 
Grating period & $\Lambda$ & 278 nm \\
Surrounding index & $n_{\circ}$ & 1 (Air) \\ 
Core index & $n_{g}$ & 1.52 \\
Substrate index & $n_{s}$ & 1.488 \\
Thermal expansion & $\alpha$ & 77 ppm $K^{-1}$ \\
\hline 
\end{tabular}
\label{tab:GeomParam}
\end{table}

\begin{table}
\centering
\caption{Measured dimensions} 
\begin{tabular}{lcc}
\hline
Name & Design & Measured \\ \hline
Core height & 2 $\mu \textrm{m}$ & 1.61 $\mu \textrm{m}$ \\
Rib height & 0.8 $\mu \textrm{m}$ & 0.79 $\mu \textrm{m}$ \\
Rib width & 20 $\mu \textrm{m}$ & 20.23 $\mu \textrm{m}$ \\
Grating period & 278 nm & 277.88 nm \\
Grating length  & -  & 12.5 mm \\
\hline
\end{tabular}
\label{tab:MeasGeomParam}
\end{table}

\begin{table}[H]
\centering
\caption{Characterised parameters}
\def\arraystretch{1.5}
\begin{tabular}{lcc}
\hline Name & Symbol & Value \\ 
\hline
Core index & $n_g$ & 1.543 ($\SI{0}{\degreeCelsius}$) \\
Substrate index & $n_s$ & 1.488 ($\SI{0}{\degreeCelsius}$ )\\
Thermo-optic coeff. Core & $N^\prime_g$ & $-1.612\cdot10^{-4}K^{-1}$ \\
Thermo-optic coeff. Substrate & $N^\prime_s$ & $-1.812\cdot10^{-4}K^{-1}$ \\
\hline
\end{tabular} 
\label{tab:CharParam}
\end{table}

\subsection{Mode excitation}\label{sect:ModTransFunc}
A final consideration is the excitation of the multiple modes in a multimode waveguide and the possible transfer of energy between them. As we discussed in Section \ref{sect:theory}\ref{sect:MMProp}, the maximum number of modes supported in a multimode waveguide is given by $p$ and the energy distribution of the modes by $\xi$ characterise $\vec{E_w}$ . Estimating how many modes are actually excited is crucial in order to compare simulations and measurements.

To model the excitation of the modes by illumination of the waveguide facet, the modal transfer function is a useful tool; the modal transfer function of a waveguide illuminated by LEDs and LDs has been studied previously \cite{Golowich.2004}. When these light sources are used with coupling lenses whose NA is larger than the NA of the waveguide, the modal transfer function can be assumed to be uniform \cite{He.1991}, so that
\begin{equation}
\Psi_j = 1.
\end{equation}
In general, however, $\Psi_j $ can take on different values. For low NA, the relationship is exponential \cite{Rittich.1985}
\begin{subequations}
	\begin{align}
	\Psi_j &= \mu e^{-\mu j}\\
	\mu &= 1
	\end{align}
\end{subequations}
and for intermediate NA we have \cite{Rittich.1985}
\begin{equation}
	\Psi_j = \begin{cases}
	1 , j \leq q\\
	0 , j > q
	\end{cases}
\end{equation}
As the modal transfer function is defined as relative energy distribution between modes, we formulate it as \cite{Olivero.2010}
\begin{equation}
\xi_j = \Psi_j / \sum_{1}^{j}\Psi_j .
\end{equation}
namely the portion of energy in a given mode normalized by the total energy of all modes.

Since the NA of the waveguide we fabricated and modeled was 0.31, an in-coupling objective with a higher NA (0.65) was used to assure uniform excitation in the subsequent experiments. We thus assumed that all possible modes may be  excited. Analytic expressions for determining the maximum number of modes are available for circular waveguides \cite{Zhao.2006} but not inverted rib waveguides.  Based on a modal analysis of the structures, mode profiles of 79 waveguide and 21 substrate modes were subsequently used in simulation.

\subsection{Spectrometer output}
The simulations were done with a resolution of 0.1 pm. This resolves each sinusodial oscillation in the single mode Bragg spectrum into 36 points. The measured spectra of the lightsource and cap sensor have a resolution of 56 pm and 35 pm. The simulated responses were rescaled so that their resolution matches with the spectrometer.

\section{Experimental Setup}
\begin{figure}[t]
\centering
\fbox{\includegraphics[width=\columnwidth,keepaspectratio=true]{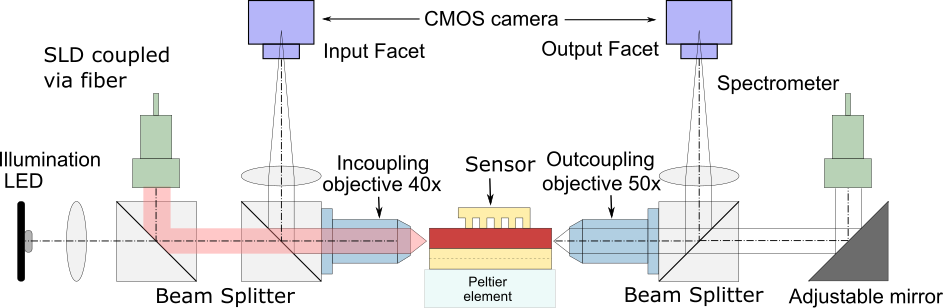}}
\caption{Experimental setup to measure the spectral response of the fabricated Bragg sensor}
\label{fig:Msetup}
\end{figure}

For experimental investigation of the fabricated Bragg sensor, a 800-860 nm super luminescent light-emitting diode (SLD-351 from SUPERLUM) is used as the light source, as shown in Figure~\ref{fig:Msetup}. The incident light passes through a beam splitter and is focused on the input facet through a RMS40X - 40X Olympus Plan Achromat objective. The light emerging from the waveguide is coupled into a spectrometer (Optical Spectrum Analyzer, AQ-6315A/-6315B, Yokogawa) through a long working distance 50x microscope objective from Mitutoyo. This output spectrum is measured from 815 to 855 nm in steps of 40 pm.

\section{Simulation \& Measurement Results}\label{sect:Meas}
We compare the simulated and measured characteristics of the different sensor configurations and derive from these an understanding of the nature of the optical transmission behavior.

\subsection{Effect of mode number on Bragg wavelength}
\begin{figure}[t]
\centering
\fbox{\includegraphics[width=0.75\columnwidth,keepaspectratio=true]{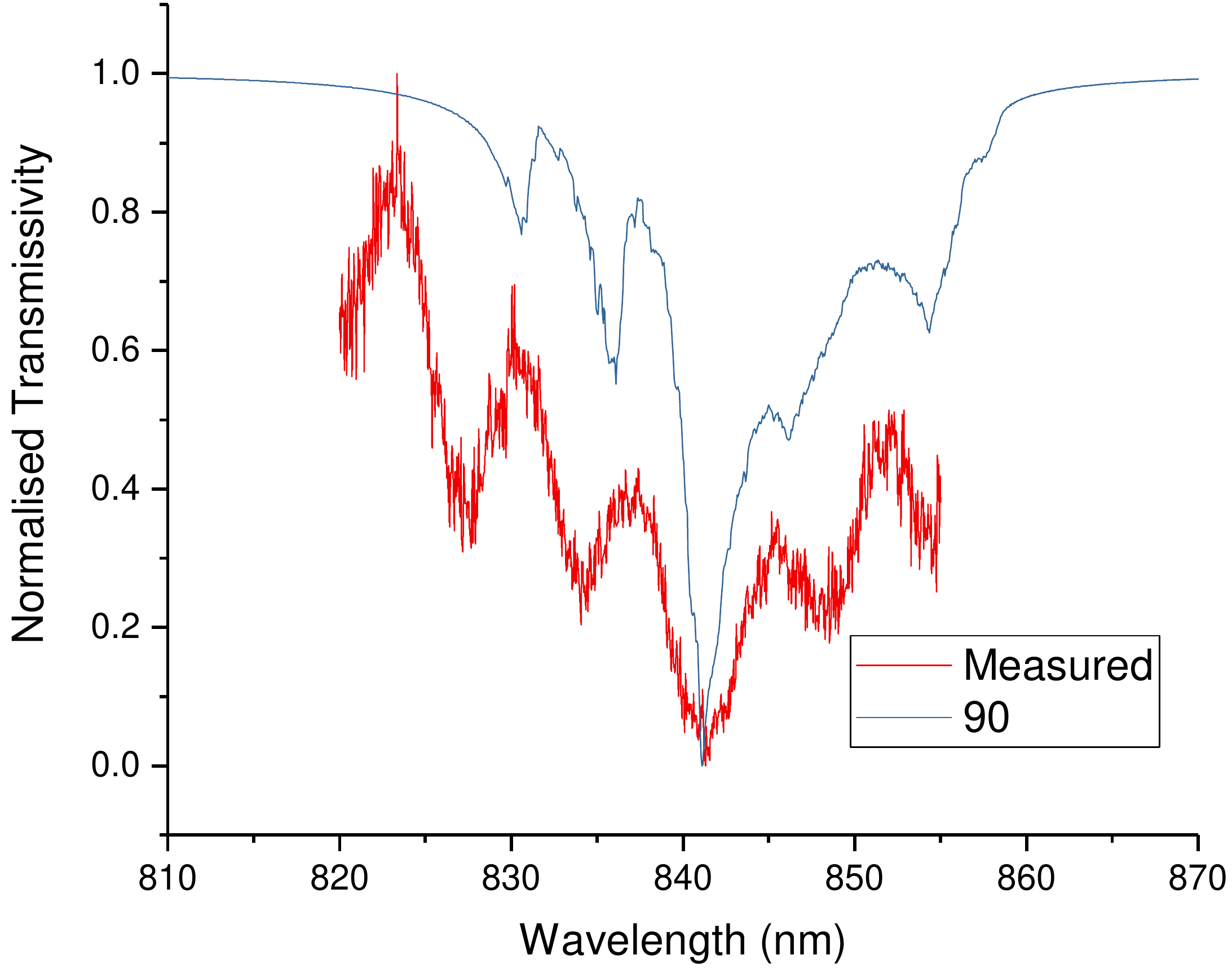}}
\caption{Measured transmissivity spectrum which exhibits multiple minima with different transmissivities. Minima to the left, visible in measured spectrum, but absent in simulated spectrum are likely due to substrate modes beyond m=99, which are not considered in the scope of the paper.}
\label{fig:Meas_trans}
\end{figure}

The measured transmitted spectrum of a cap sensor (recall Figure~\ref{fig:CapSensor}) on a multi-mode waveguide is shown in Figure~\ref{fig:Meas_trans}. Multiple dips in the transmission can be seen, in contrast to the characteristic of a single mode Bragg grating, which shows only a single dip \cite{Kocabas.2006}. The simulated spectrum, also shown in the figure, explains the origin of the multiple minima: they result from the spectral overlap of multiple modes. The simulated results are offset to the right by $5.5nm$ on account of manufacturing tolerances.

\begin{figure}[t]
\centering
\fbox{\includegraphics[width=0.45\columnwidth,keepaspectratio=true]{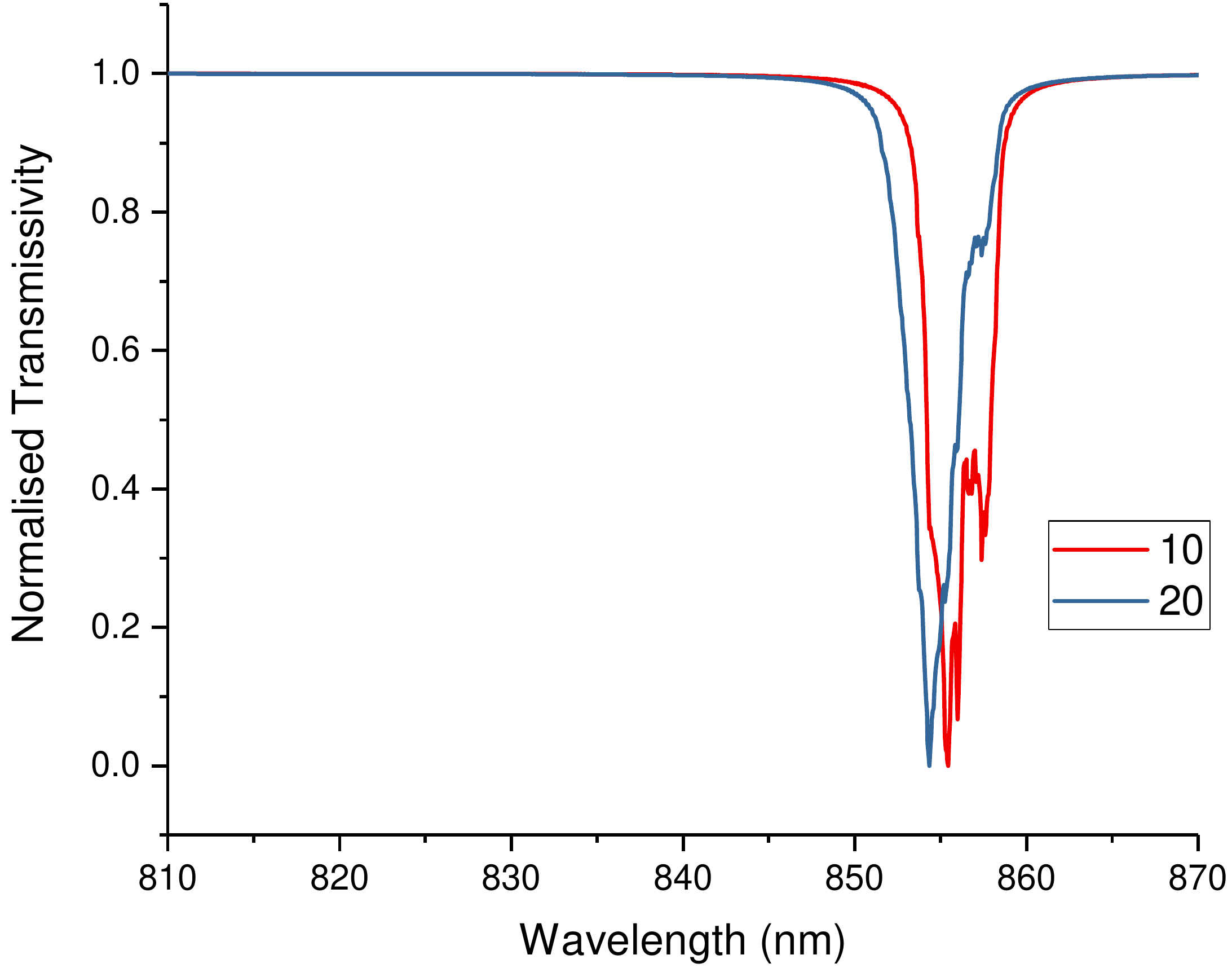}}
\fbox{\includegraphics[width=0.45\columnwidth,keepaspectratio=true]{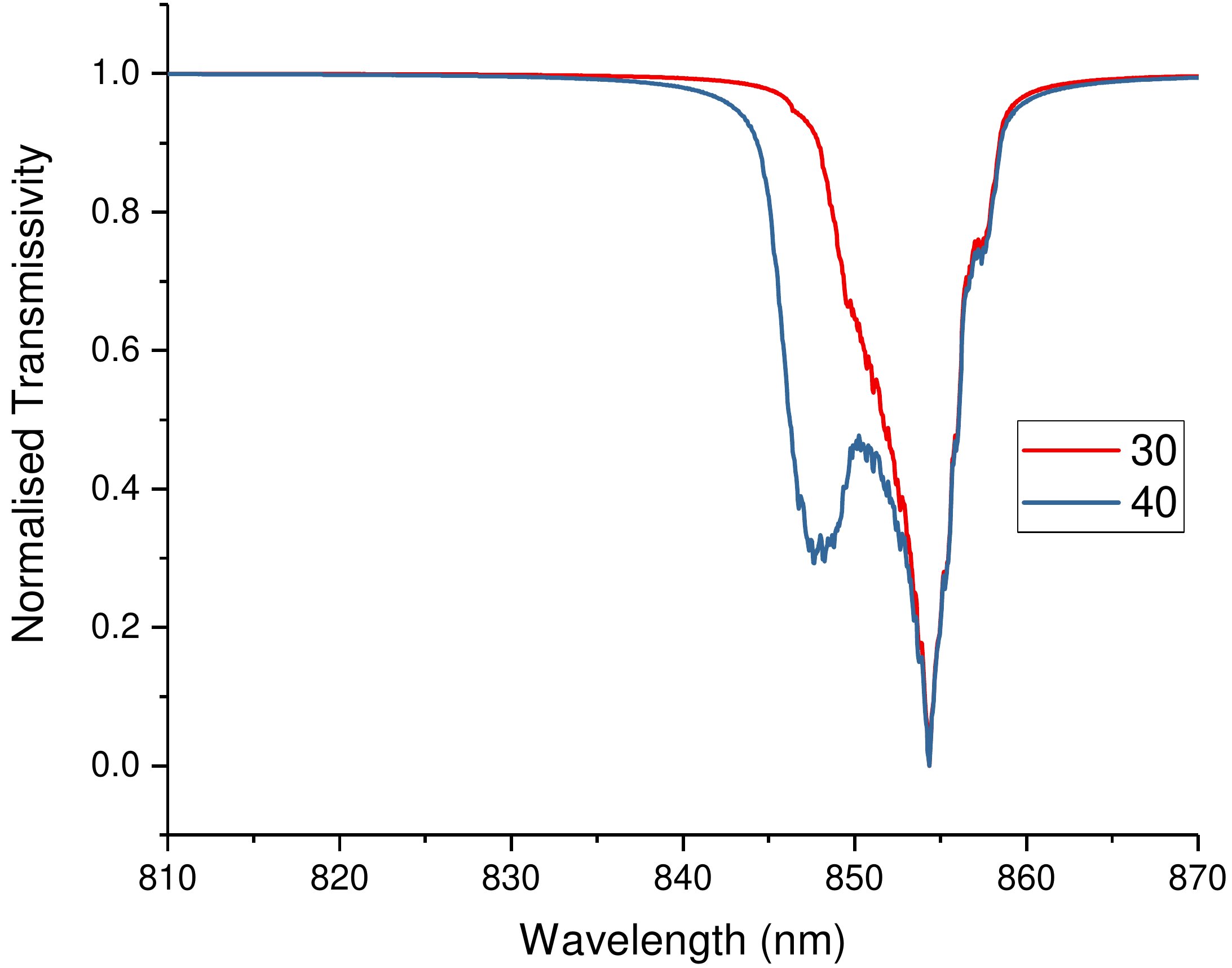}}
\fbox{\includegraphics[width=0.45\columnwidth,keepaspectratio=true]{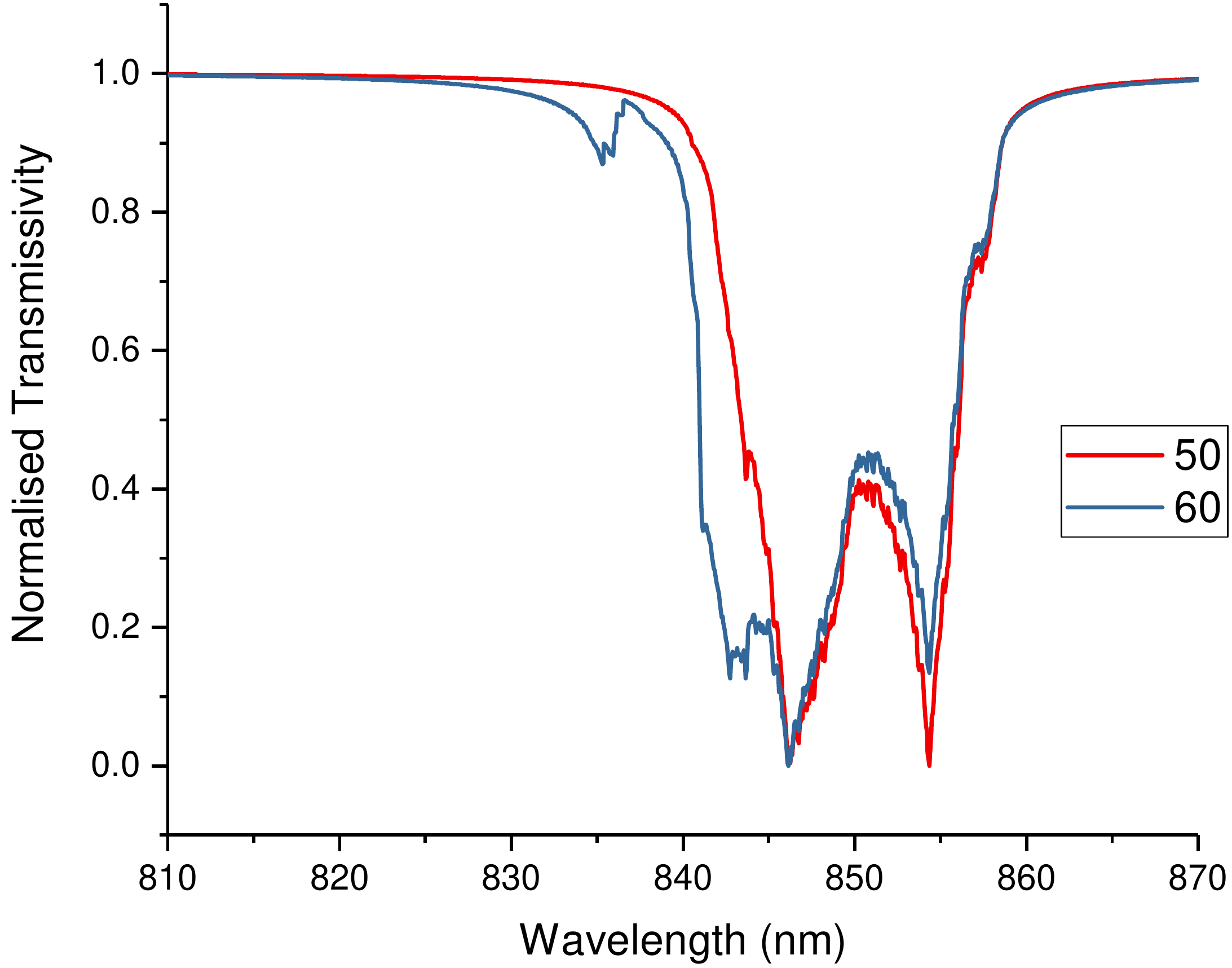}}
\fbox{\includegraphics[width=0.45\columnwidth,keepaspectratio=true]{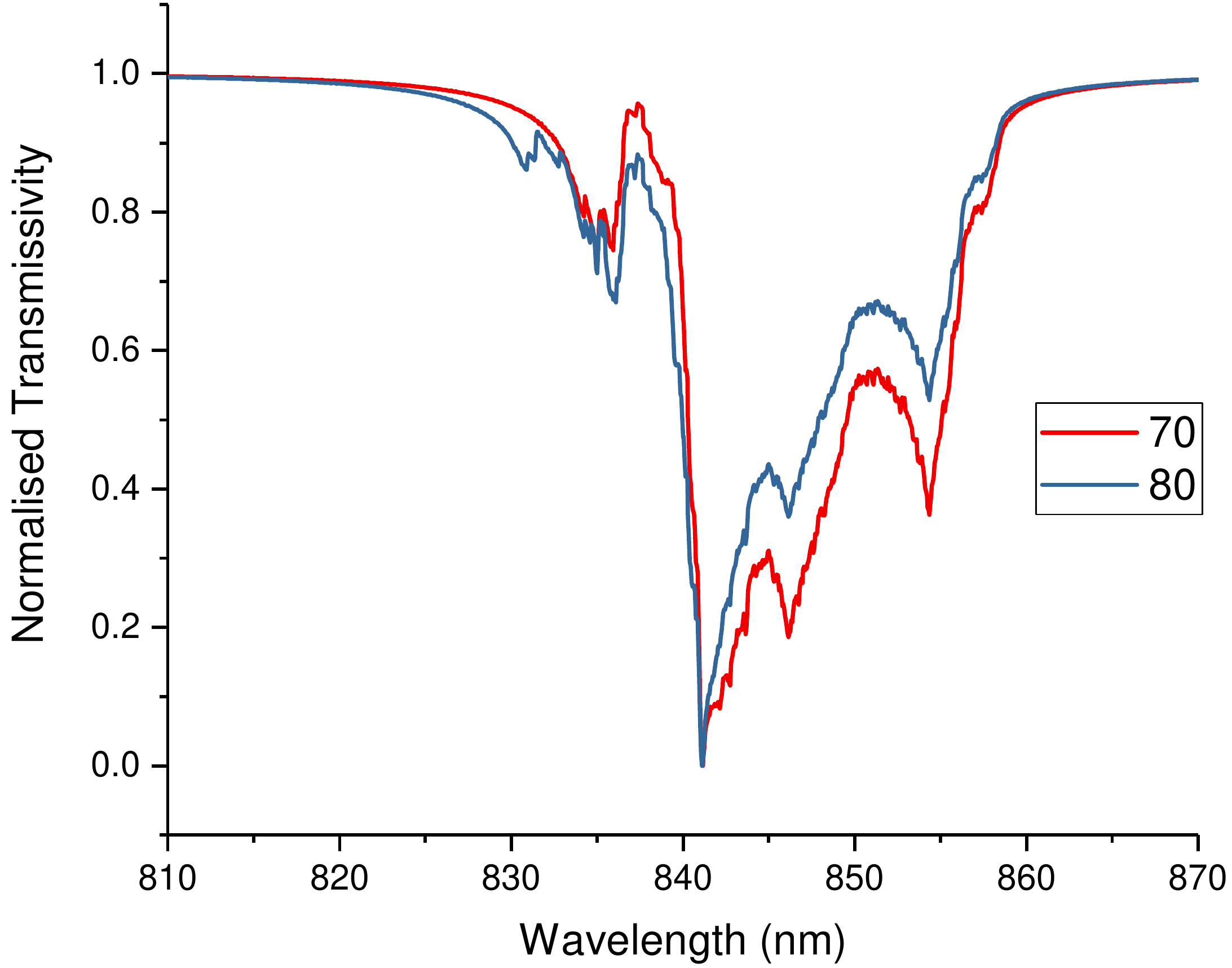}}
\fbox{\includegraphics[width=0.45\columnwidth,keepaspectratio=true]{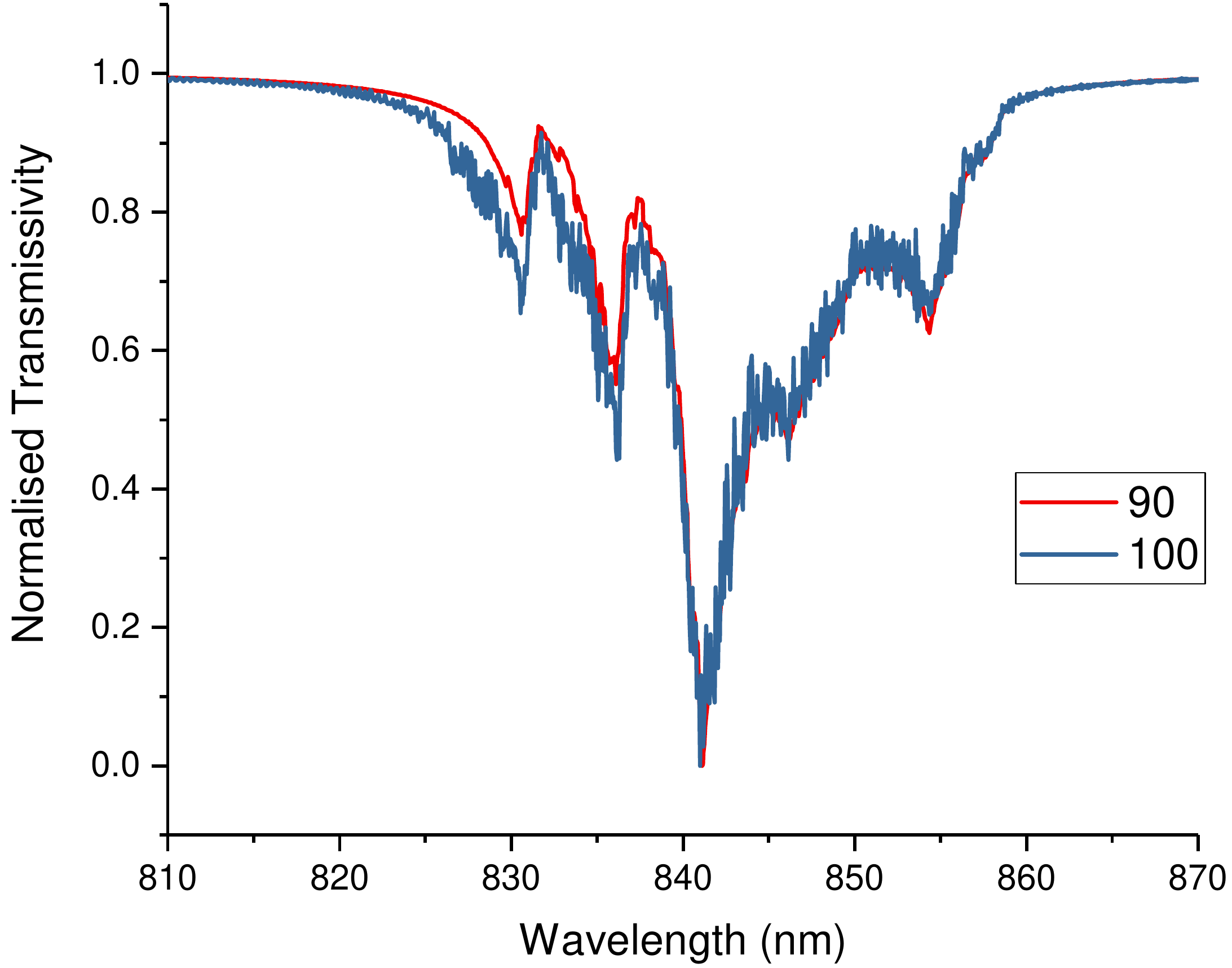}}
\caption{Simulated spectra vs number of modes: additional minima appear as the number of modes considered is increased.}
\label{fig:BW_Spectrometer}
\end{figure}

In contrast with a single mode Bragg grating, where the Bragg wavelength is solely determined by the grating period and core refractive index, the number of excited modes and the modal transfer function also affect the Bragg wavelength in a multimode surface grating. In simulation, when the number of modes considered is increased, the output spectrum significantly changes, as shown in Figure~\ref{fig:BW_Spectrometer}. The dips to the right of the spectrum are caused by waveguide modes and dips to the left, by substrate modes. The results show how important it is to estimate the number of propagating modes correctly, if the output spectrum is to be accurately modeled.

\subsection{Effect of grating length on transmissivity}
\begin{figure}[t]
\centering
\fbox{\includegraphics[width=0.75\columnwidth,keepaspectratio=true]{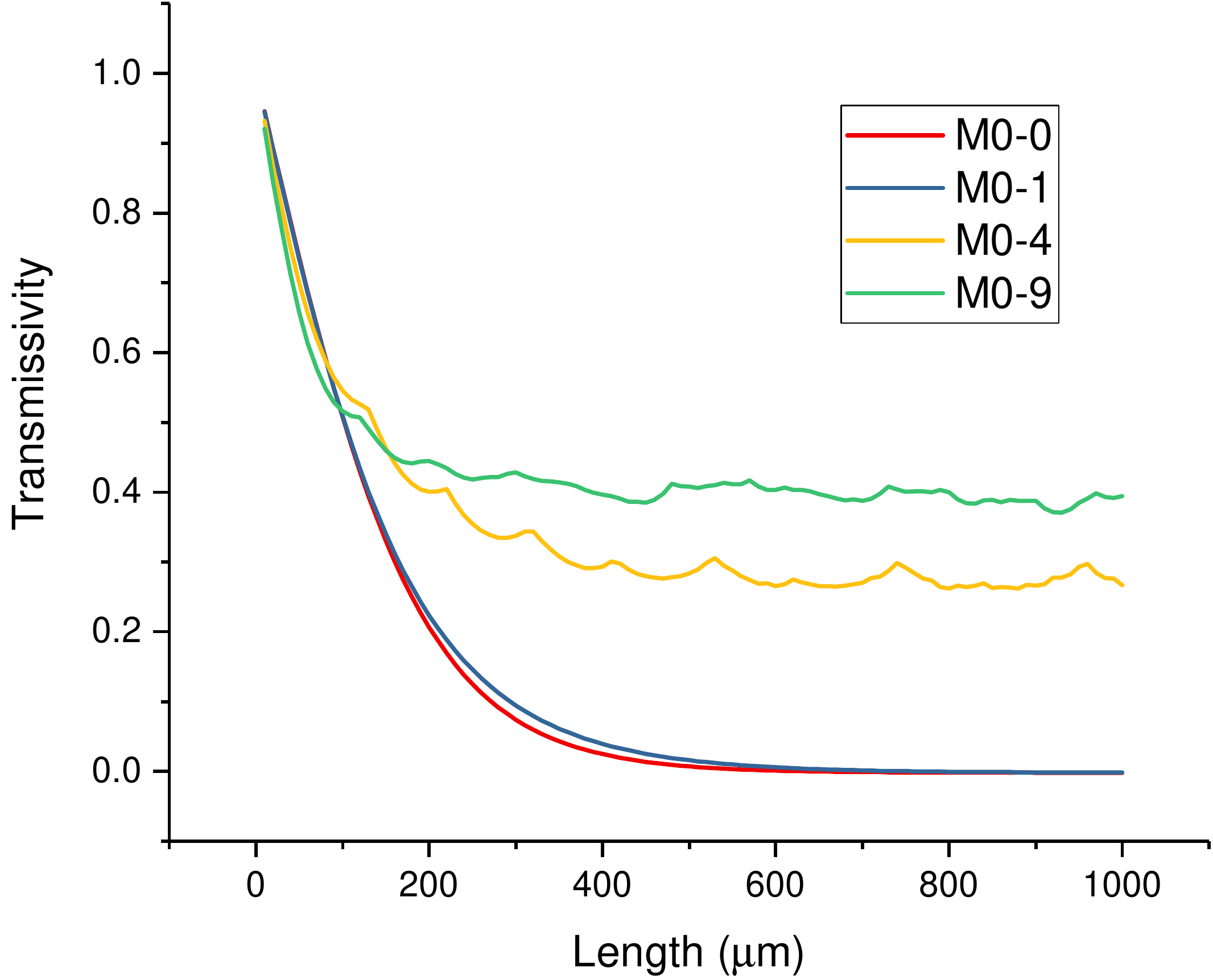}}
\caption{Evolution of transmissivity with length for multimode gratings}
\label{fig:EffectofLength}
\end{figure}

The transmissivity vs grating length was analyzed as a function of the number of modes. Whereas the single mode transmissivity for a sufficiently long grating always goes to zero \cite{Erdogan.1997}, the measured multimode transmissivity does not reach zero, as seen in Figure~\ref{fig:EffectofLength}. Although light in the vicinity of Bragg wavelength is reflected completely within each mode, the $\lambda_b$ of each mode is different. Hence light which is reflected from lower modes propagates unhindered in higher modes and vice versa. Since the spectrometer measures the total intensity at each wavelength irrespective of which mode the light came from,  the dips in the simulated and measured spectrum for most multimode Bragg gratings, will not reach zero.

\subsection{Cap vs Core Bragg sensor - spectrum}
\begin{figure}[t]
\centering
\fbox{\includegraphics[width=0.75\columnwidth,keepaspectratio=true]{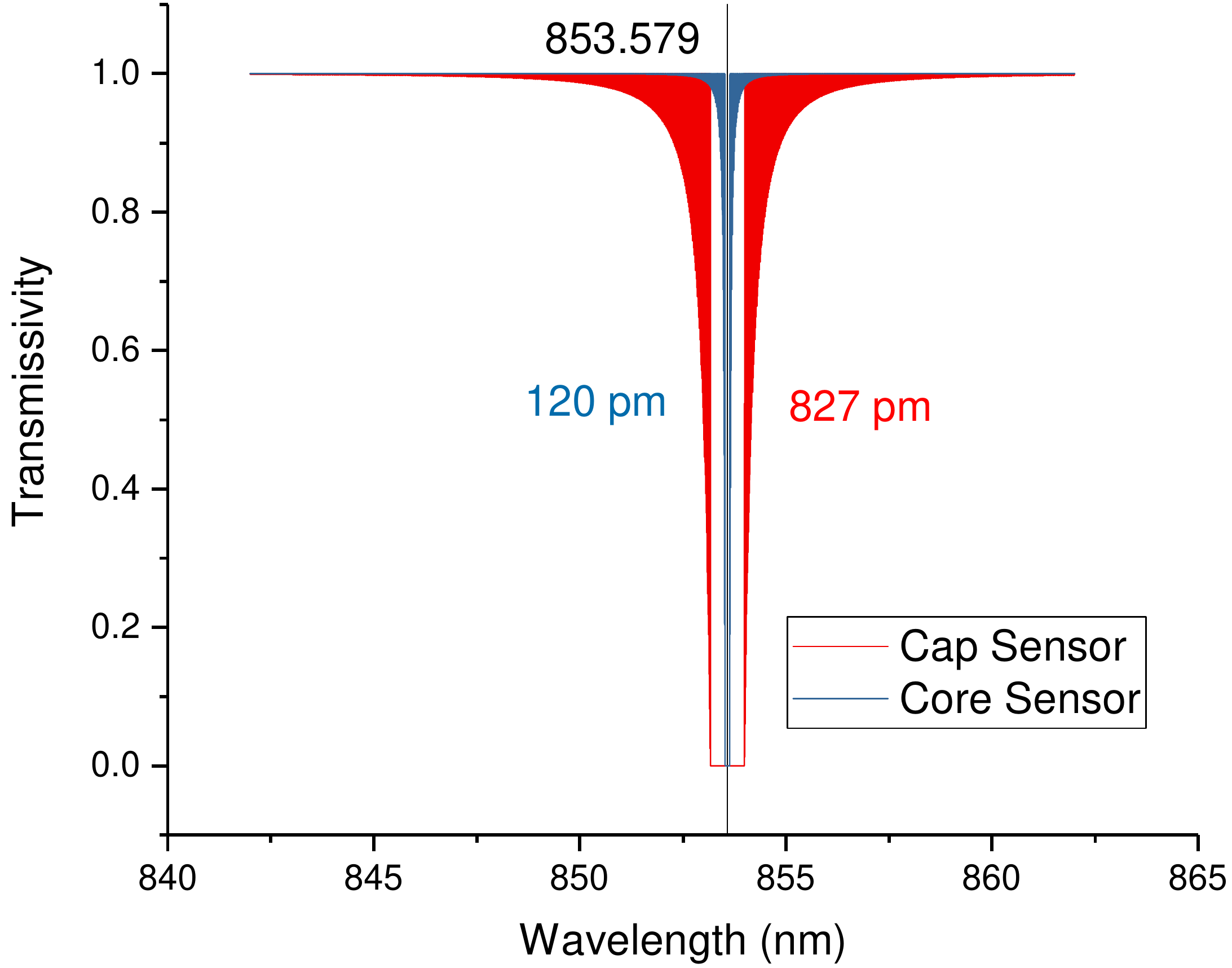}}
\caption{Modes of a cap sensor are wider than those of a core sensor. Single mode}
\label{fig:SM_CapvsCore}
\end{figure}

\begin{figure}[t]
\centering
\fbox{\includegraphics[width=0.75\columnwidth,keepaspectratio=true]{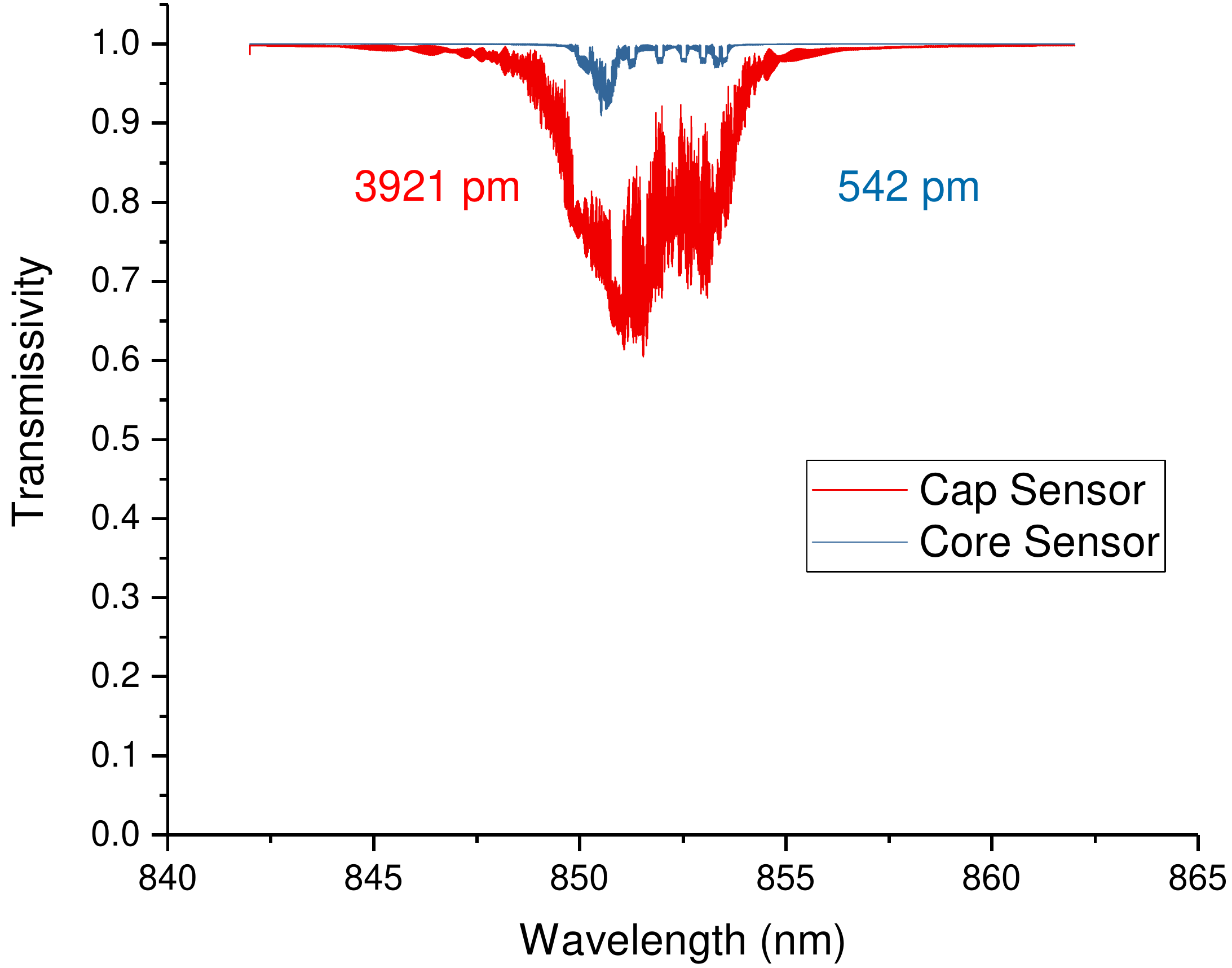}}
\caption{Modes of a cap sensor are wider than those of a core sensor. Multimode}
\label{fig:MM_CapvsCore}
\end{figure}

Finally, we compare the cap and core sensor configurations (recall Figures \ref{fig:CapSensor} and \ref{fig:CoreSensor}). The transmissivity of the cap sensor is lower than that of the core sensor, as seen in Figures \ref{fig:SM_CapvsCore} and \ref{fig:MM_CapvsCore}. The origin of this difference is that the overlap between modes is higher in a cap sensor, and the FWHM of each mode is \SI{800}{\pico\meter} as compared to \SI{120}{\pico\meter} for a core sensor, although their Bragg wavelengths are the same \SI{853.578}{\nano\meter}. The higher FWHM of a cap sensor is explained by its higher index difference between the grating and the surrounding material ($\delta n_{cap} =  0.488$) than is the case for a core sensor ($ \delta n_{core} = 0.055$). As a result, the high transmissivity of the core sensor will make its peak wavelength cumbersome to detect and the cap sensor is more suitable for detection of the Bragg wavelength in a sensor application.

\subsection{Cap vs Core Bragg sensor - sensitivity}
\begin{figure}[t]
\centering
\fbox{\includegraphics[width=0.75\columnwidth,keepaspectratio=true]{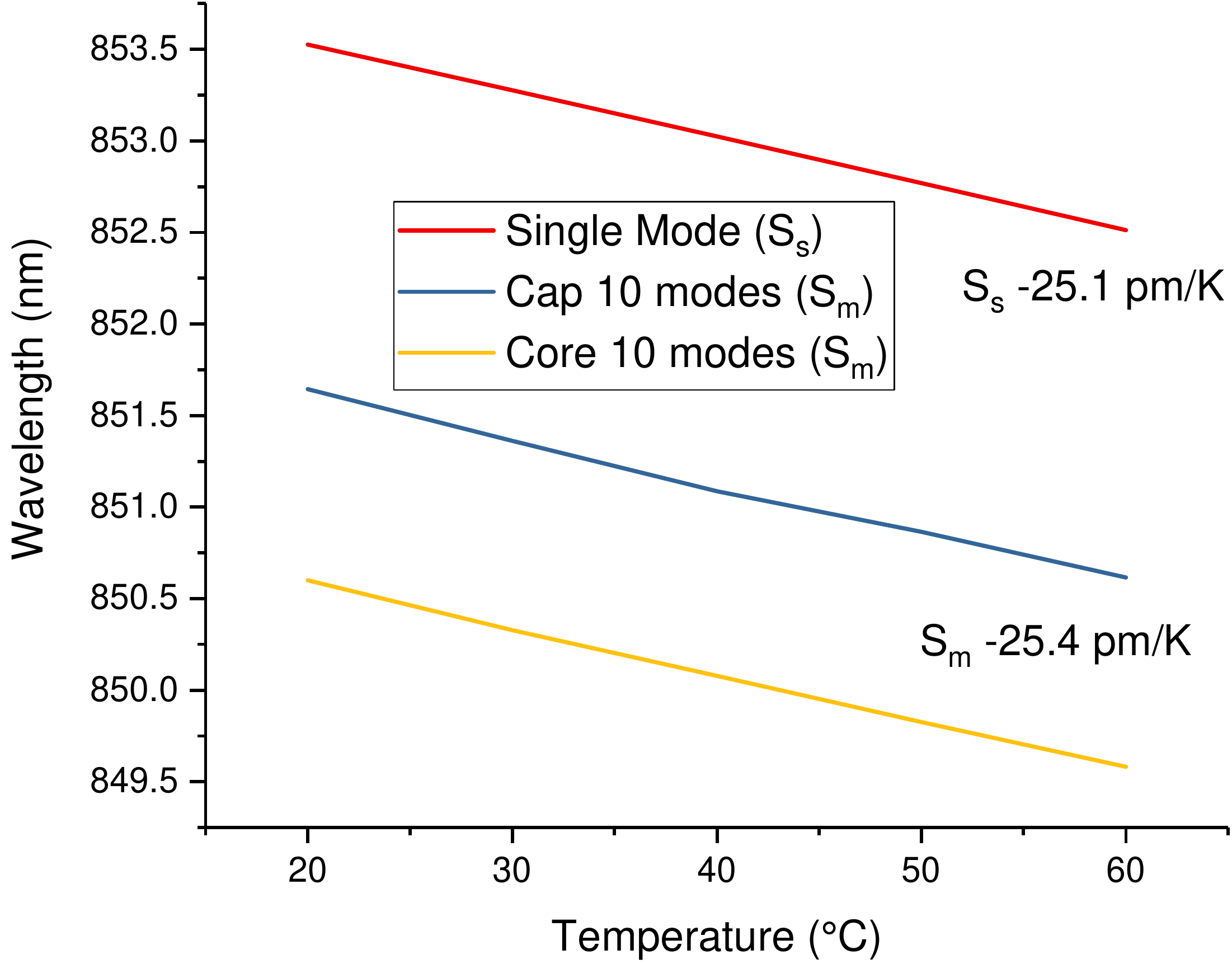}}
\caption{Simulated sensitivity of multimode and single mode Bragg sensors are nearly identical.}
\label{fig:Sensitivity}
\end{figure}

The measured temperature sensitivity (-306 pm/K) of the multimode RBG and its simulated single mode sensitivity ($S_s = -25 pm/K$ ) have been compared and published previously \cite{Sherman.2016}. As seen in Figure~\ref{fig:Sensitivity}, it was found that the sensitivity of the cap sensor ($S_m$) remains the same in the single mode and multimode domains, at a value of -25.4 pm/K. The difference in measured and simulated sensitivities is possibly due to effect of polymer cross linking at the UGS45E-PMMA interface. An unforeseen assumption made implicitly in the calculations was that exposure to UV light creates UGS45E without interaction with PMMA. However, ethylene glycol dimehyl acrylate and PMMA have been shown to react with each other under UV exposure \cite{Isaure.2004, Kingsbury.2011, Lesturgeon.1999}. As this reaction can directly affect the refractive index of the core, it is the likely cause of the higher sensitivity.

\subsection{Discussion}\label{sect:Discussion}
As we have seen, numerical simulation enables an understanding of the spectral behavior of multimode Bragg sensors. Some differences between the measured and simulated spectrum still remain: for example, the dips in the measured spectrum are wider, possibly due to shifts in polarization during propagation.

The redistribution of power due to scattering causes the multimode transmissivity to decrease more in experiment than in the simulations. The net effect is the coupling of light with a wavelength $\lambda$ from modes that are not reflected by the grating into modes at wavelengths which are reflected. Thus as more light is removed from the transmitted beam, the corresponding dips in that portion of the spectrum will become wider. An additional experimental effect is that the grating period varies along the $z$ axis due to manufacturing imperfections, which further augments redistribution of power among modes, making the spectral dips wider.

Upon examining the light source spectrum and the reference waveguide spectrum, we realised that the loss coefficients have a noticeable wavelength-dependent distribution, causing the spectrum to shift.

Finally, in the scope of this paper, apodisation functions were assumed to be zero, which effectively means the grating period is constant. Non uniform Bragg gratings allow the spectral response to be tailored by introduction of phase shifts, side lobe suppression and dispersion compensation. The derivation can be applied for non uniform gratings by modifying the description of coupling coefficient, as proposed by Erdogan \cite{Erdogan.1997}.

\section{Conclusion}
It could be shown that the reflected wavelength of multimode Bragg-reflector waveguides is affected by the energy distribution among the modes and the number of supported modes. An equation for multimode transmissivity usable for multimode waveguides of circular and non circular cross sections, with full or partial index modulation, has been numerically derived. For an inverted rib waveguide with a uniform modal transfer function and 90 modes, the simulated results correlated well with the experimentally measured spectrum.

These simulations give valuable insights into features of multimode Bragg grating spectra. For example, we have seen that there are multiple spectral dips due to overlap of modes and that the number of dips in the transmitted spectrum increases with increasing number of excited modes. In addition, for surface Bragg gratings, the intra-modal coupling coefficients dominate over inter-modal coupling coefficients; as several planar waveguides have this feature, eigen-decomposition can used to speed up their spectral simulations.

\section{Appendices}

\subsection{Appendix A - Calculation of coupling coefficients}

The Poynting vector of a mode has the following expression\cite{Floris.2010},
\begin{equation}
S_j = \iint \vec{e_{t,j}} \times \vec{h_{t,j}}\cdot dx \cdot dy
\end{equation}

The mode profiles as calculated using Beam Propogation Method generated 2D plots of normalised magnitude of electric field $\vec{E}$ as functions of $x$ and $y$ . In BPM, a 3D infinite wave is modelled as a 2D plane wave with boundary conditions. In such cases, one can estimate the magnitude of the Poynting vector as:

\begin{equation}
S_j = \frac{2}{\eta_\circ}\iint |\vec{E(x,y)}|^2 \cdot dx dy
\end{equation}

$\eta_\circ$ is the resistance in vacuum as is equal to 377 ohms.

The transverse coupling coefficient $\kappa^t_{kj}$ can be expressed in terms of permittivity profile $\epsilon$, transverse electric field profile of the mode $\vec{e^t_j}$ and Poynting vector as:

\begin{equation}\label{eq:kappa}
\kappa_{j} = \frac{\omega }{2}
\frac{\iint(\epsilon - \epsilon_\circ) \vec{e_{t,j}}^2 \cdot dx \cdot dy }
{S_j} 
\end{equation}

\subsection{Appendix B - Refractive index side profile}\label{sect:appendixB}

\begin{figure}[H]
	\begin{minipage}{0.45\columnwidth}
		\begingroup%
		\makeatletter%
		\setlength{\unitlength}{\textwidth}
		\makeatother%
		\begin{picture}(1,1.00000023)%
		\put(0,0){\fbox{\includegraphics[width=\unitlength,page=1]{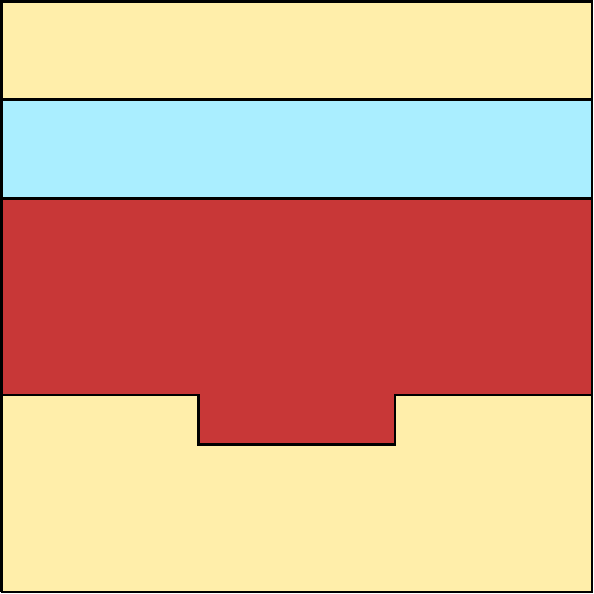}}}%
		\put(0.49863078,0.90134013){\color[rgb]{0,0,0}\makebox(0,0)[b]{\smash{$n_{s}$}}}%
		\put(0.49594497,0.73545359){\color[rgb]{0,0,0}\makebox(0,0)[b]{\smash{$n_{\circ}$}}}%
		\put(0.499289,0.47324794){\color[rgb]{0,0,0}\makebox(0,0)[b]{\smash{$n_{g}$}}}%
		\put(0.49797249,0.08862791){\color[rgb]{0,0,0}\makebox(0,0)[b]{\smash{$n_{s}$}}}%
		\put(0,0){\fbox{\includegraphics[width=\unitlength,page=2]{Pics_Sensor_CrossSectionView_06_Cap.pdf}}}%
		\put(0.08677833,0.20131232){\color[rgb]{0,0,0}\makebox(0,0)[b]{\smash{y}}}%
		\put(0.1979683,0.06796838){\color[rgb]{0,0,0}\makebox(0,0)[b]{\smash{x}}}%
		\put(0.04088154,0.06763177){\color[rgb]{0,0,0}\makebox(0,0)[b]{\smash{z}}}%
		\end{picture}
		\endgroup%
		\caption{Cap Bragg Sensor- Front View}
		\label{fig:Cross_Cap}
	\end{minipage}\hfill
	\begin{minipage}{.45\columnwidth}
		\begingroup%
		\makeatletter%
		\setlength{\unitlength}{\textwidth}
		\makeatother%
		\begin{picture}(1,1.00000023)%
		\put(0,0){\fbox{\includegraphics[width=\unitlength,page=1]{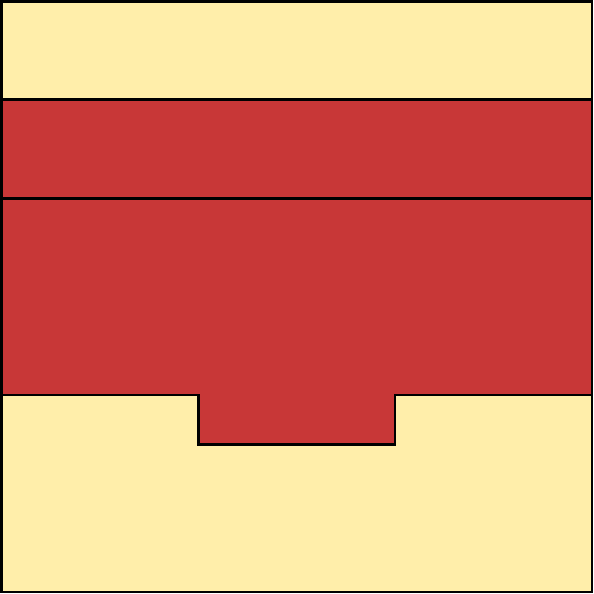}}}%
		\put(0.50597716,0.89465193){\color[rgb]{0,0,0}\makebox(0,0)[b]{\smash{$n_{s}$}}}%
		\put(0.50460793,0.09131393){\color[rgb]{0,0,0}\makebox(0,0)[b]{\smash{$n_{s}$}}}%
		\put(0.50663545,0.50000017){\color[rgb]{0,0,0}\makebox(0,0)[b]{\smash{$n_{g}$}}}%
		\put(0.4991777,0.72518354){\color[rgb]{0,0,0}\makebox(0,0)[b]{\smash{$n_{g}$ }}}%
		\put(0,0){\fbox{\includegraphics[width=\unitlength,page=2]{Pics_Sensor_CrossSectionView_04_Core.pdf}}}%
		\put(0.08677833,0.20131232){\color[rgb]{0,0,0}\makebox(0,0)[b]{\smash{y}}}%
		\put(0.1979683,0.06796838){\color[rgb]{0,0,0}\makebox(0,0)[b]{\smash{x}}}%
		\put(0.04088154,0.06763177){\color[rgb]{0,0,0}\makebox(0,0)[b]{\smash{z}}}%
		\end{picture}%
		\endgroup%
		\caption{Core Bragg Sensor- Front view}
		\label{fig:Cross_Core}
	\end{minipage}
\end{figure}

\begin{figure}[H]
	\centering
	\begingroup%
	\makeatletter%
	\setlength{\unitlength}{\columnwidth}
	\makeatother%
	\begin{picture}(1,0.37070602)%
	\put(0,0){\fbox{\includegraphics[width=\unitlength,page=1]{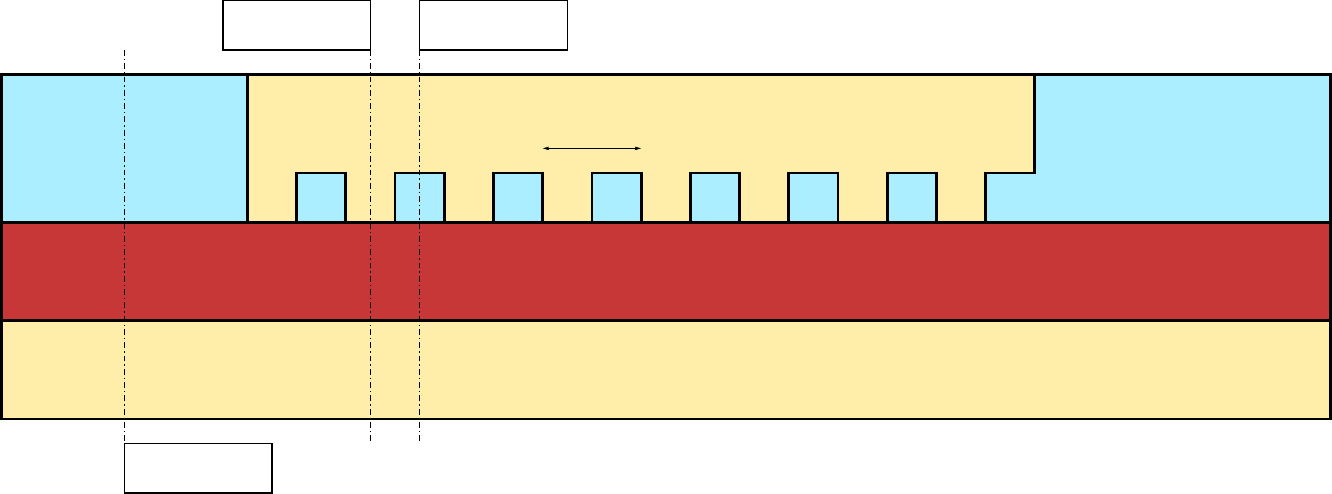}}}%
	\put(0.44489239,0.26657631){\color[rgb]{0,0,0}\makebox(0,0)[b]{\smash{$\Lambda$}}}%
	\put(0.37154996,0.34080044){\color[rgb]{0,0,0}\makebox(0,0)[b]{\smash{Fig \ref{fig:Cross_Cap}}}}%
	\put(0.15322085,0.00997091){\color[rgb]{0,0,0}\makebox(0,0)[b]{\smash{Fig \ref{fig:CrossWav}}}}%
	\put(0.22183902,0.34118841){\color[rgb]{0,0,0}\makebox(0,0)[b]{\smash{Fig \ref{fig:Geometric}}}}%
	\put(0,0){\fbox{\includegraphics[width=\unitlength,page=2]{Pics_Sensor_SideView_03_Cap.pdf}}}%
	\put(0.03941375,0.28047373){\color[rgb]{0,0,0}\makebox(0,0)[b]{\smash{y}}}%
	\put(0.01892457,0.22198655){\color[rgb]{0,0,0}\makebox(0,0)[b]{\smash{x}}}%
	\put(0.08937031,0.22034641){\color[rgb]{0,0,0}\makebox(0,0)[b]{\smash{z}}}%
	\end{picture}%
	\endgroup%
	\caption{Cap Bragg Sensor - Side view}
	\label{fig:Side_Cap}
\end{figure}

Figure \ref{fig:Side_Cap} shows the side view of a cap sensor while Figure \ref{fig:Cross_Cap} shows the cross-sectional view of its grating.

\begin{figure}[H]
	\centering
	\begingroup%
	\makeatletter%
	\setlength{\unitlength}{\columnwidth}
	\makeatother%
	\begin{picture}(1,0.37070657)%
	\put(0,0){\fbox{\includegraphics[width=\unitlength,page=1]{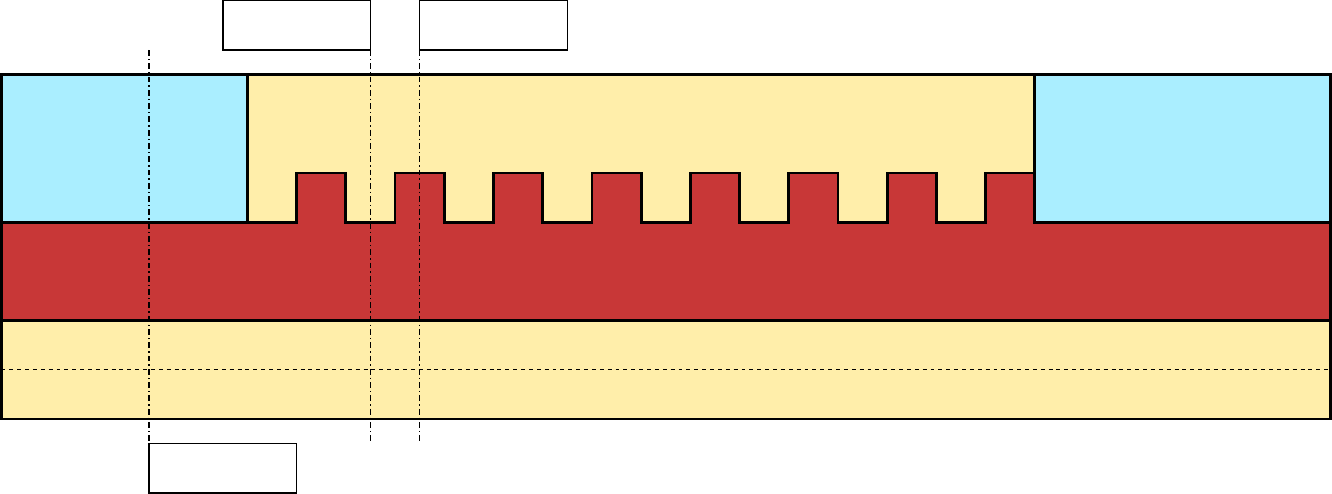}}}%
	\put(0.22487399,0.34250579){\color[rgb]{0,0,0}\makebox(0,0)[b]{\smash{Fig \ref{fig:Geometric}}}}%
	\put(0.37240261,0.34087202){\color[rgb]{0,0,0}\makebox(0,0)[b]{\smash{Fig \ref{fig:Cross_Core}}}}%
	\put(0.16626421,0.00797057){\color[rgb]{0,0,0}\makebox(0,0)[b]{\smash{Fig \ref{fig:CrossWav}}}}%
	\put(0,0){\fbox{\includegraphics[width=\unitlength,page=2]{Pics_Sensor_SideView_02_Core.pdf}}}%
	\put(0.03941378,0.29156427){\color[rgb]{0,0,0}\makebox(0,0)[b]{\smash{y}}}%
	\put(0.01892458,0.23307704){\color[rgb]{0,0,0}\makebox(0,0)[b]{\smash{x}}}%
	\put(0.08937038,0.2314369){\color[rgb]{0,0,0}\makebox(0,0)[b]{\smash{z}}}%
	\put(0.49053446,0.26349858){\color[rgb]{0,0,0}\makebox(0,0)[b]{\smash{L}}}%
	\put(0.49051116,0.15691196){\color[rgb]{0,0,0}\makebox(0,0)[b]{\smash{$N^\prime_g$}}}%
	\put(0.48967259,0.08566424){\color[rgb]{0,0,0}\makebox(0,0)[b]{\smash{ $N^\prime_s$}}}%
	\end{picture}%
	\endgroup%
	\caption{Core Bragg Sensor - Side view}
	\label{fig:Side_Core}
\end{figure}

Figure \ref{fig:Side_Core} shows the side view of a core sensor while Figure \ref{fig:Cross_Core} shows the cross-sectional view of its grating.

\subsection{Appendix C - FEM parameters}\label{sect:appendixC}
Parameters affecting the numerical accuracy with which CMT \& BPM are computed using the finite element method (FEM).
\begin{table}[H]
	\centering
	\caption{FEM settings} 
	\begin{tabular}{lc}
		\hline
		Name & Settings \\ 
		\hline 
		Grid size & 20 nm \\
		Grid type & Uniform \\
		Slices per grating & 4 \\
		Wavelength & 852 nm\\
		Resolution & 0.1 pm \\
		Tolerance & $10^{-4}$ \\
		Number of modes & 100 \\
		Wavelength range & $\SI{782}{\nano\meter}$ - $\SI{882}{\nano\meter}$ \\
		Precision & Single float \\
		\hline 
	\end{tabular} 
	\label{tab:SoftParam}
\end{table}

\bibliography{citations}
\ifthenelse{\equal{\journalref}{ol}}{%
\clearpage
\bibliographyfullrefs{citations}
}{}

\end{document}